\documentclass{article}

\usepackage[hidelinks]{hyperref}
\usepackage{arxiv}

\usepackage[utf8]{inputenc} 
\usepackage[T1]{fontenc}    
\usepackage[hidelinks]{hyperref}       
\usepackage{url}            
\usepackage{booktabs}       
\usepackage{amsfonts}       
\usepackage{nicefrac}       
\usepackage{microtype}      
\usepackage{graphicx}
\usepackage{epstopdf, epsfig}
\usepackage{amsmath}
\usepackage{graphicx}
\usepackage{graphics}
\usepackage{listings}

\usepackage{natbib}	
\usepackage{gensymb}
\usepackage{comment}
\usepackage{enumitem}
\setitemize{noitemsep,topsep=0pt,parsep=0pt,partopsep=0pt,leftmargin=*}
\usepackage{amssymb}

\usepackage{lineno}
\usepackage{setspace}
\usepackage{footnote}

\usepackage{tabularx}

\usepackage{nopageno}

\usepackage{fancyhdr}
\usepackage{subcaption}

\usepackage{xcolor}
\usepackage{multirow}

\title{Turbulent kinetic energy dissipation from colliding ice floes }
\author{
  Trygve K. Løken$^{a,}$\footnotemark , Aleksey Marchenko$^{b}$, Thea J. Ellevold$^{a}$, Jean Rabault$^{c}$ and Atle Jensen$^{a}$
}

\begin{document}
\maketitle
 
\footnotetext{E-mail address and phone number for corresponding: trygve.loken@gmail.com (+47) 94885767 (T.K. Løken)\\ 
E-mail addresses: alekseym@unis.no (A. Marchenko), theajel@math.uio.no (T.J. Ellevold), \\ jean.rblt@gmail.com (J. Rabault) and atlej@math.uio.no (A. Jensen)}

\noindent $^{a}$ Department of Mathematics, University of Oslo, Oslo, Norway \\
$^{b}$ University Centre in Svalbard, Longyearbyen, Norway \\
$^{c}$ Norwegian Meteorological Institute, Oslo, Norway \\


\begin{abstract}

Increased knowledge about wave attenuation processes in sea ice, and hence atmosphere-wave-ice-ocean energy transfer, is necessary to improve sea ice dynamics models used for climate modeling and offshore applications. The aim of this study is to generate such much needed data by investigating colliding ice floes dynamics in a large-scale experiment and directly measuring and quantifying the turbulent kinetic energy (TKE). The field work was carried out at Van Mijen Fjord on Svalbard, where a 3$\times$4~m ice floe was sawed out in the fast ice. Wave motion was simulated by pulling the ice floe back and forth in an oscillatory manner in a 4$\times$6~m pool, using two electrical winches. Ice floe motion was measured with a range meter and accelerometers, and the water turbulence was measured acoustically with an acoustic Doppler current profiler and optically with a remotely operated vehicle and bubbles as tracers. TKE frequency spectra were found to contain an inertial subrange where energy was cascading at a rate proportional to the -5/3 power law. The TKE dissipation rate was found to decrease exponentially with depth. The total TKE dissipation rate was estimated by assuming that turbulence was induced over an area corresponding to the surface of the floe. The results suggest that approximately 37\% and 8\% of the input power from the winches was dissipated in turbulence and absorbed in the collisions, respectively, which experimentally confirms that energy dissipation by induced turbulent water motion is an important mechanism for colliding ice floe fields.

\end{abstract}

\section{Introduction} \label{sec:Introduction}

A decline in the Arctic ice cover has been observed over the past decades \citep{feltham2015arctic}, which has allowed for more human activities in the region, such as shipping, tourism and exploitation of natural resources \citep{smith2013new}. Better predictions of sea ice hazards are necessary to ensure safe operations in the Marginal Ice Zone (MIZ), which is the transition between the land fast ice or dense pack ice and the open ocean, consisting of a distribution of ice floes in size at various concentrations. On the one hand, the retreating ice cover leads to larger areas of open water in the Arctic where more energetic waves are generated due to the increased fetch, which in turn enhance ice break up processes \citep{thomson2014swell}. On the other hand, experimental studies have shown that waves are exponentially damped in the MIZ \citep{squire1980direct,wadhams1988attenuation}, meaning that the presence of the MIZ mitigates ice cover break up. This interplay illustrates that wave-ice interactions, which are coupled in a nonlinear manner, are key mechanisms for the Arctic. There is uncertainty associated with the dominating source of wave energy dissipation by sea ice. Increased knowledge about these physical processes, and hence atmosphere-wave-ice-ocean energy transfer, is necessary to improve sea ice dynamics models used for wave forecasts and climate modeling. 

Several phenomena are known to attenuate waves in an ice floe field, such as wave scattering or directional spreading and viscous dissipation in the boundary layer below the ice due to shear flow or wake formation caused by a relative velocity between the water and the ice \citep{wadhams1975airborne,liu1988wave,marchenko2019wave}. Scattering, which contributes to wave decay due to energy reflection and spreading, is known to be of importance in open fields of floes where the floe diameter is of the same order as the ocean wavelength \citep{squire1995ocean,squire2018fresh}. Ice floe interactions can lead to wave energy dissipation through different mechanisms and are of relevance in denser fields. Collisions between neighboring ice floes can for example cause momentum transfer and energy absorption during the impulse \citep{shen1998wave,herman2018wave,li2018laboratory}. Immersed collision between small spheres and a wall have been investigated experimentally (e.g. \cite{joseph2001particle}) and numerically (e.g. \cite{li2012contact}), and the loss of kinetic energy has been attributed to inelasticity during impact and viscous stresses in the liquid. Although these works elaborate on the mechanisms of energy dissipation in collisions, they may not adequately describe the processes occurring in the MIZ due to differences in scaling and geometry. \cite{rabault2019experiments} showed from wave tank experiments that colliding chunks of grease ice can generate turbulence that injects eddy viscosity in the water, which leads to enhanced energy dissipation. However, scaling problems in for example Reynolds number are inevitable in laboratory experiments, which pleads for performing full-scale measurements outside of the laboratory, see e.g. \cite{rabault2019experiments} for a discussion on the topic. 

Mathematical models have been developed to describe wave attenuation in the MIZ, e.g. the viscoelastic model of \cite{zhao2018three} and the viscous models of \cite{sutherland2019two} and \cite{marchenko2019wave}. \cite{sutherland2019two} leave freedom of interpretation of the effective viscosity while \cite{marchenko2019wave} associate the effective viscosity with the eddy viscosity. These models rely on physical parameters, e.g. the effective viscosity, that may be adjusted through curve-fitting exercises to match experimental data, although they may lack direct proof of which phenomena that are of importance. By contrast, direct observations on the full scale can describe in detail the mechanisms occurring. There are few in situ observations of the water kinematics around interacting ice floes because the harsh conditions make field work challenging. \cite{voermans2019wave} managed to measure under-ice turbulence in pancake and frazil ice generated from the relative velocity between the ice and the orbital wave motion and suggested that turbulence dissipation caused wave attenuation. \cite{marchenko2015characteristics} measured turbulence under continuous drift ice and found that the main source of under-ice turbulence was associated with water motion relative to the ice caused by tidal current and wind drift of the ice. However, the effect of turbulent dissipation around larger interacting ice floes, typically found in the Greenland Sea and Arctic MIZ, has not been previously confirmed experimentally. 


In this study, direct observations of the turbulent kinetic energy (TKE) dissipation rate in the immediate vicinity of a colliding full-scale ice floe are presented for the first time. A high level of control over the floe motion and the surrounding water velocity was obtained from an extensive instrumentation, which would have been extremely challenging to deploy in the dynamic and hazardous environment of the MIZ. Hence, an outdoor laboratory on an ice-covered fjord was installed as a compromise between realistic scale and high level of control. An ice floe was cut out from the land fast ice. Ice floes respond in surge when acted upon by wave trains entering the MIZ and collisions between adjacent floes occur since they are moving out of phase with each other \citep{squire1995ocean}. Since there was negligible wave energy at the selected site, the ice floe was towed back and forth to simulate wave motion. The experimental setup was similar to the one of \cite{marchenko2021field}, who measured turbulent properties with an acoustic Doppler velocimeter (ADV). The novelty of the current experiment is the use of an acoustic Doppler current profiler (ADCP), which enabled the authors to estimate the TKE dissipation rate on several locations to quantify the importance of turbulence induced from collisions and shear flow, and the use of a remotely operated vehicle (ROV) which, together with a bubble seeding system, allowed the authors to observe the general structure of flow motion under the ice. Energy dissipated in collisions was determined from high resolution accelerometer data. The extensive instrumentation allowed for control of input energy rates and thus estimates of a floe energy balance.   

The paper is organized in the following manner. Experimental setup, data acquisition and processing methods are described in Section~\ref{sec:Data_methods}. Section~\ref{sec:theoretical_background} contains a mathematical description of the problem. The results in Section~\ref{sec:Results} are presented as an energy budget where the rate of energy input is compared with the rate of dissipation. Finally, a discussion on the accuracy and implications of the results follows in Section~\ref{sec:Discussion}, and the concluding remarks are given in Section~\ref{sec:Conclusions}.

\section{Data and methods} \label{sec:Data_methods}



\begin{figure}
	\centering
	\subcaptionbox{Map indicating experimental site.\label{fig:map}}
		{\includegraphics[width=.35\linewidth, height=.34\linewidth]{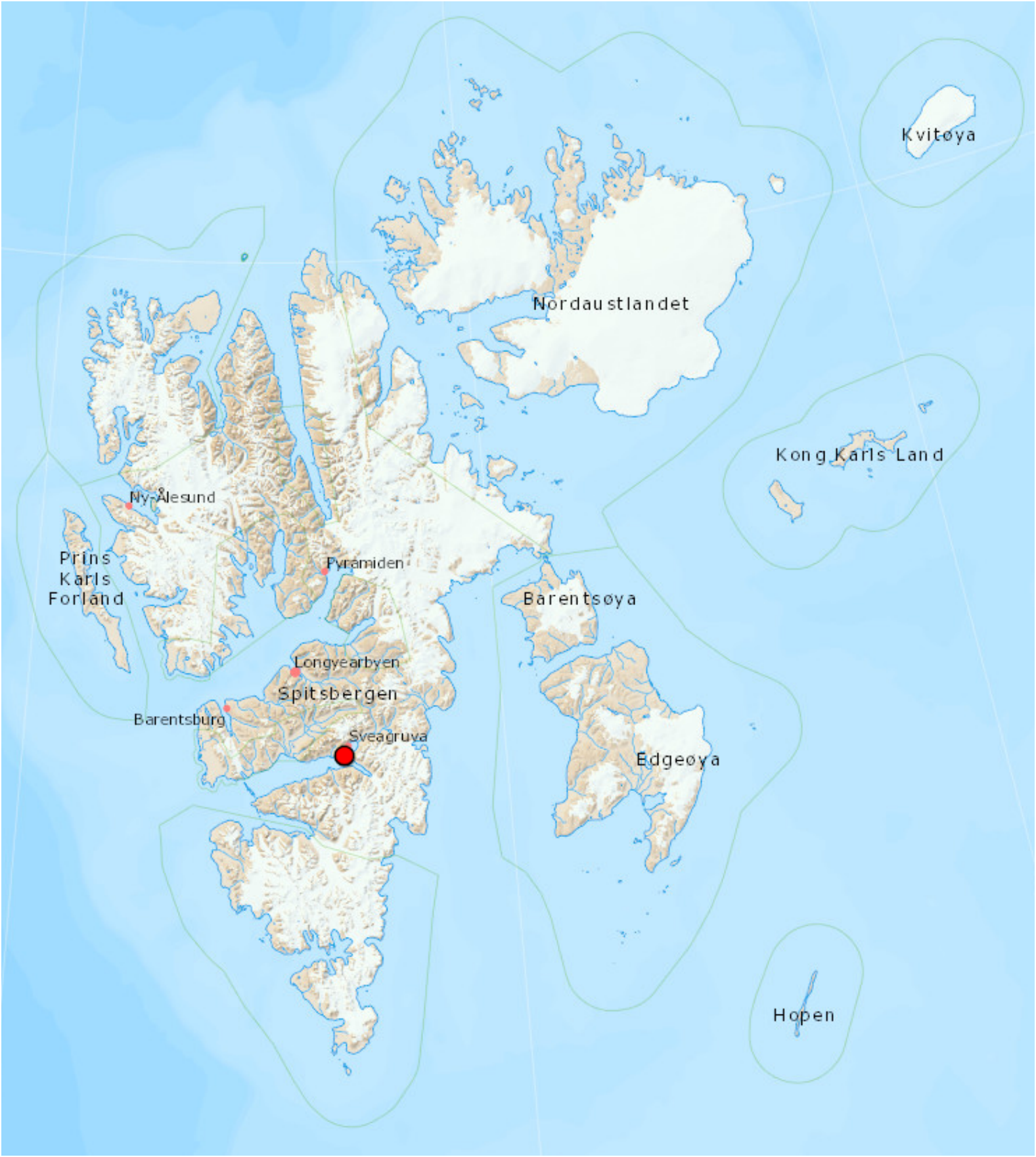}} 
	\subcaptionbox{Preparation of field laboratory.\label{fig:preparation}}
		{\includegraphics[width=.45\linewidth, height=.34\linewidth]{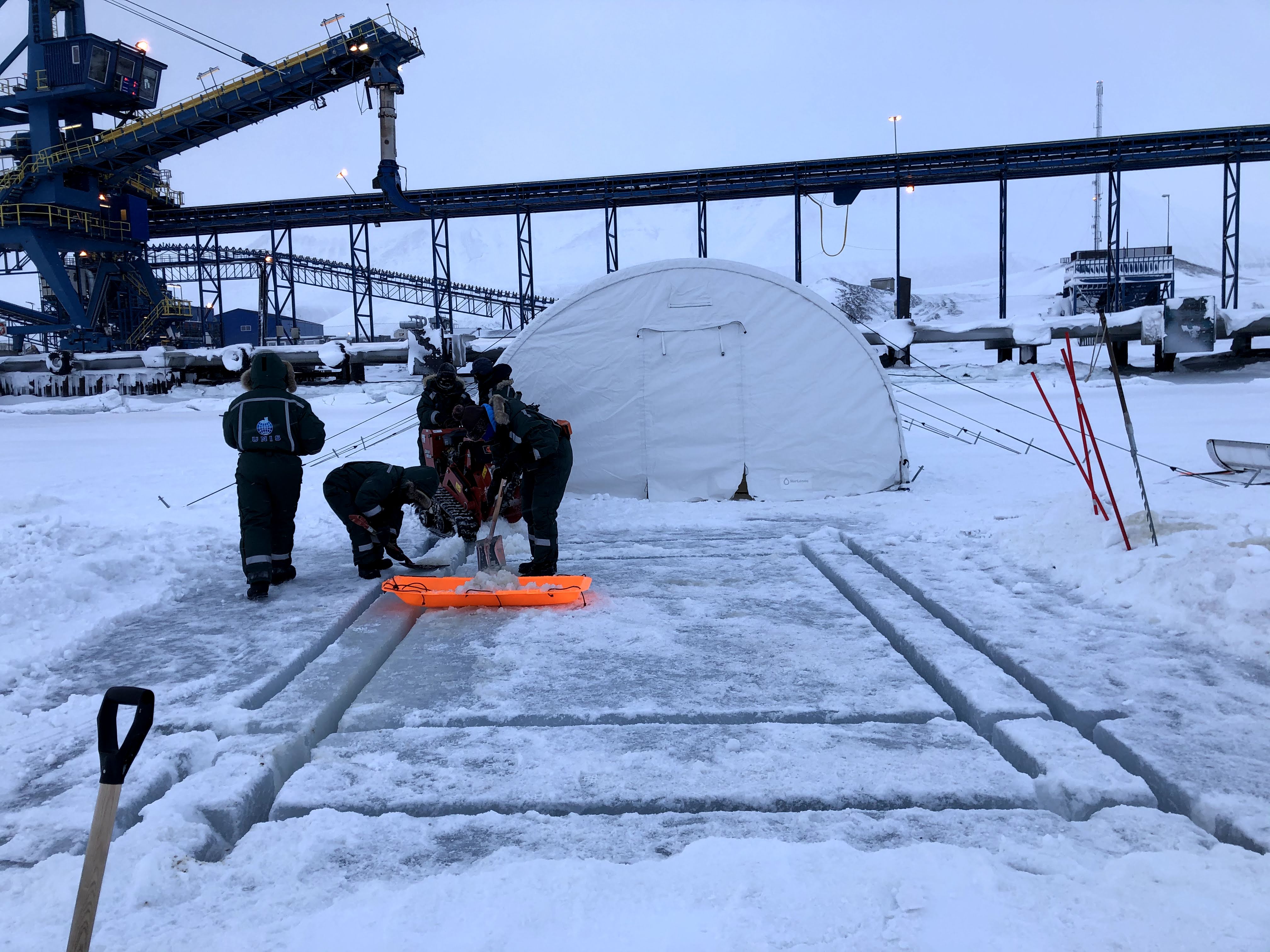}}
	\caption{Location and preparation of the field laboratory. a) A map of the area where the experimental work was carried out (red dot indicates the location). Source: \cite{TopoSvalbard}. b) The working process of cutting the ice. The frame between the outer and inner rectangle was removed to create a floating ice floe in a pool. Afterwards, the inflatable tent in the background was placed over the pool for weather protection. }\label{fig:map_preparation}
\end{figure}

The field work was carried out next to the harbor in the Svea Bay on Svalbard on March 3-12, 2020. The location is indicated with a red dot in figure~\ref{fig:map} and the geographical coordinates were 77.86$^{\circ}$N, 16.65$^{\circ}$E. Svea Bay is part of the Van Mijen Fjord which was covered with land fast ice at the time of the field campaign. An ice floe was made at the selected site where the ice thickness was approximately 1~m. Figure~\ref{fig:preparation} shows an outer and an inner rectangle measuring 6$\times$4~m and 4$\times$3~m, respectively, which were cut through the sea ice by means of a walk-behind chain trencher and hand saws. The ice between the two rectangles were broken into manageable blocks and removed with chains and hoists installed on a quadpod lifting rig, resulting in a floating ice floe in a pool. A 10$\times$6~m inflatable tent was placed over the pool for weather protection and equipped as a field laboratory.   

A coordinate system, shown in figure~\ref{fig:setup}, was defined with the $(x,y,z)$-axis to be aligned horizontally in the axial and transverse direction of the pool and vertically in upward direction, respectively. The $x$-axis was oriented with an angle $\alpha=28^{\circ}$ counterclockwise from the magnetic north. Hence, the short ends of the pool were defined as the north and south ends. The origin was defined as $x=0$ at the pool south end, $y=0$ at the pool center and $z=0$ at the bottom of the ice. The coordinate system included in figure~\ref{fig:setup_AD} is displaced along the $z$-axis to the top side of the ice for increased readability. The floe dimensions $L_f$, $W_f$ and $H_f$ in the $x$, $y$ and $z$-directions were 4, 3 and 1~m, respectively. 

At the location of the field laboratory, there was negligible wave energy. Therefore, two electrical winches were used to tow the ice floe back and forth in an oscillatory manner in the $x$-direction to simulate wave induced motion. One period of oscillation, i.e. the floe motion back and forth, will be referred to as a \textit{cycle}. The winches were mounted to the fast ice by means of ice screws, one on each short end, approximately 3~m from the pool at $y=0$. A wooden frame was attached to the floe with ice screws and the winch wires were coupled to the frame via a polyester silk rope as illustrated in figure~\ref{fig:setup_AD} to distribute the winch load over a large area of the floe surface. The winches were alternating in pulling and slacking and were manually actuated by two persons.  


\subsection{Instrumentation} \label{subsec:Instrumentation}

Six experiments, summarized in table~\ref{Table:runs}, are included in this paper. The only variables that were changed between the experiments were the number of cycles, the location and cell configuration of the ADCP and the inclusion of a load cell and accelerometers. All other parameters, such as the towing speed and the duration of the cycles, were kept approximately constant in all the experiments. The similar setup was used several times to investigate the repeatability of the experiments. 

\begin{table}[h]
\centering 
\begin{tabular}{c c c c c c}  
\toprule
\multirow{2}{*}{Experiment} & \multicolumn{2}{c}{Cycles [N]} & \multicolumn{2}{c}{ADCP} & \multirow{2}{*}{Load cell} \\[0.5ex]
\cmidrule(lr){2-3} 
\cmidrule(lr){4-5} 
{} & Total & ADV & Cells [N] & Position [m] & {} \\ [0.5ex]
\midrule
1 & 15 & 15 & 95 & 0.5 & - \\[0.5ex]
2 & 14 & 14 & 39 & 0.5 & - \\[0.5ex]
3 & 11 & 5 & 95 & 0.5 & \checkmark \\[0.5ex]
4 & 28 & 20 & 39 & 0.25 & - \\[0.5ex]
5 & 7 & 5 & 39 & 0.25 & - \\[0.5ex]
6 & 8 & - & 39 & Ice floe & - \\[0.5ex]
\bottomrule
\
\end{tabular}
\caption{Experimental details. The ADCP position is distance from the pool edge. }
\label{Table:runs} 
\end{table}     

\begin{figure}
	\centering
	\subcaptionbox{Seen from the south end of the pool.\label{fig:setup_AD}}
		{\includegraphics[width=.493\linewidth]{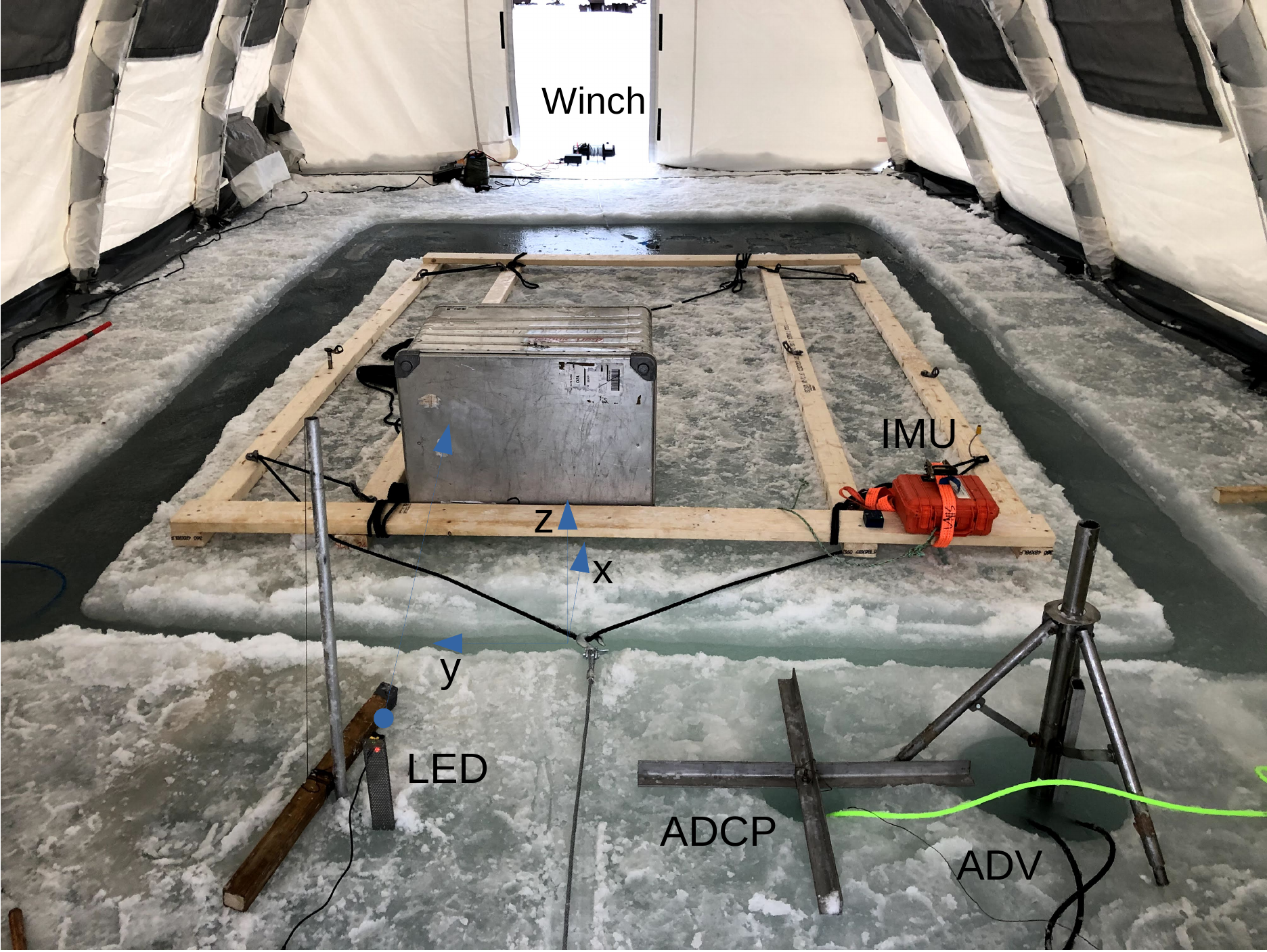}} 
	\subcaptionbox{Seen from above.\label{fig:setup_sketch}}
		{\includegraphics[width=.408\linewidth]{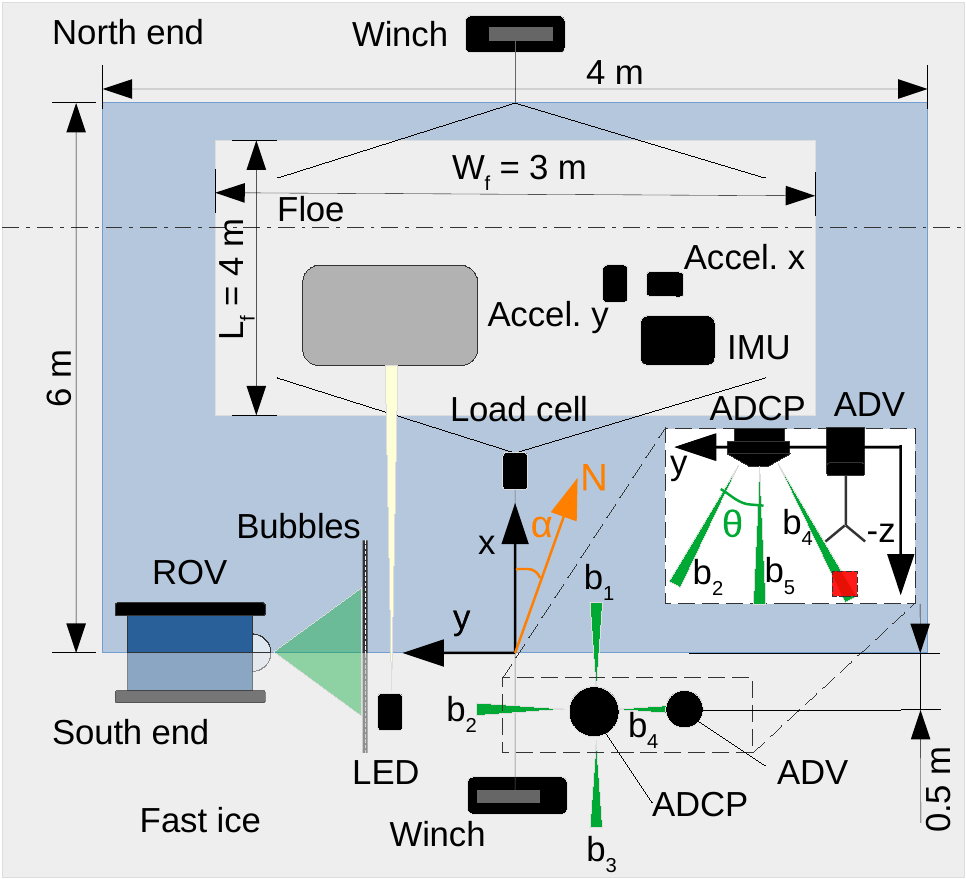}}
	\caption{Experimental setup. a) Photo of the setup seen from the south end of the pool. The defined coordinate system is indicated (although displaced from the origin along the $z$-axis to the top side of the ice for the illustrative purpose). The load cell and the uniaxial accelerometers were not installed during this particular experiment. b) Schematic of the setup in the $xy$-plane. Magnetic north (N) is indicated. The inset sketch shows the acoustic instruments in the $yz$-plane and the measurement volume of the ADV is marked with a red square, which coincides with a part of the ADCP $\mathbf{b_4}$.}\label{fig:setup}
\end{figure}
  
\begin{figure}
  \begin{center}
    \includegraphics[width=.98\textwidth]{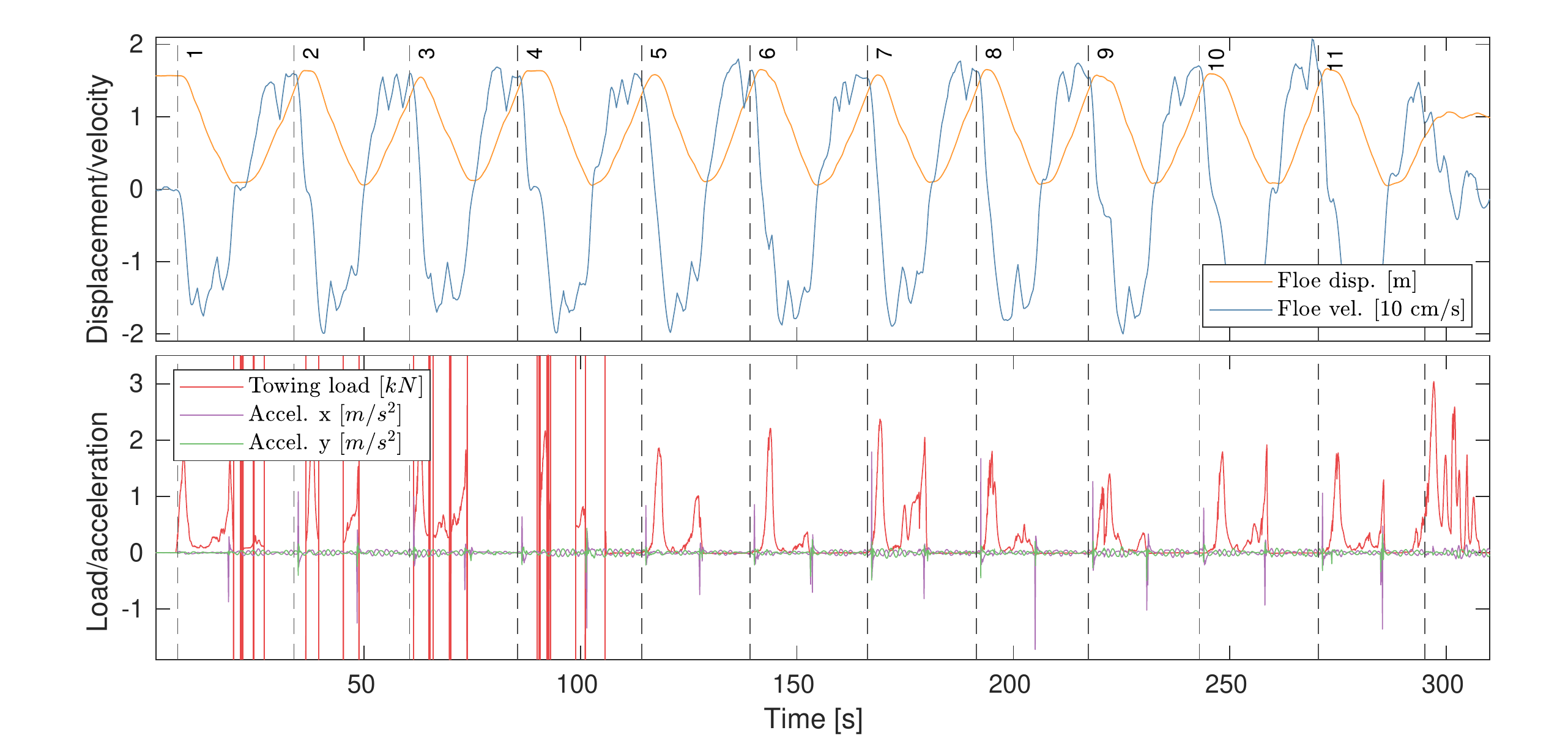} 
    \caption{\label{fig:timeseries_evo} Time series from Exp.~3 where the cycles are marked with numbers and separated with vertical dashed lines. Upper panel: range meter data with smoothed displacement and velocity of the ice floe in the $x$-direction. Lower panel: load cell and uniaxial accelerometers. Note that Cycles~2-4 contain severe load cell dropouts.}
  \end{center}
\end{figure}

Several sensors and instruments were installed on the south end of the pool, as shown in figure~\ref{fig:setup}, to measure the ice floe and water motion. An evo60 LED (light-emitting diode) range meter was pointing towards a large box placed on the floe, which provided time series of the floe surge, i.e. displacement in the $x$-direction. The sample frequency of the range meter was approximately 125~Hz and the raw data were smoothed with a moving average over 200 data points. The computer that was used to control the range meter was synchronized with Internet time each day. The ice floe velocity in the $x$-direction was found from the smoothed position with a central difference scheme. An example of a time series from Experiment~3, where the floe undergoes 11 full cycles, is displayed in the upper panel of figure~\ref{fig:timeseries_evo}. The floe was displaced approximately 1.7~m and the maximum towing velocity was quite constant and about 0.15~m/s in each direction. The oscillating period in ice floe surge, i.e. the duration of one cycle, was around 26~s.


During Exp.~3, a load cell (PCM BD-ST-620) was mounted in the coupling between the winch wire and the polyester silk rope. Only one load cell was available, and it was installed at the south end of the pool, which means that it measured towing force applied by the winch on the ice floe in the $-x$-direction. In the same experiment, two uniaxial accelerometers (Bruel and Kjær, DeltaTron Type 8344) suitable for collision measurements, were mounted on the floe, one aligned with the $-x$-direction and the other with the $y$-direction, as seen in figure~\ref{fig:setup_sketch}. The sampling frequency of the load cell and the accelerometers was 5~kHz and the signals were smoothed with a Savitzky-Golay filter over 500 data points. An example of a time series from Exp.~3 is displayed in the lower panel of figure~\ref{fig:timeseries_evo}, where the accelerometer data contain two high-amplitude events per cycle, corresponding to collision with the fast ice, and low-amplitude oscillations with a period around 2~s in between, possibly associated with surface waves in the pool. The load cell and the two accelerometers were connected to the same data acquisition unit and were therefore synchronized. However, the computer used to control the instruments was not synchronized with Internet time. In the post-processing, it was necessary to synchronize the range meter and load cell data in time, as there was a mismatch between the computer clocks. Table~\ref{Table:instruments} lists the instruments that were synchronized in the post-processing, their sampling frequencies and smoothing parameters. Details on the synchronization scheme for the instruments marked with diamonds in table~\ref{Table:instruments} can be found in Appendix~A. 


\begin{table}[h]
\centering 
\begin{tabular}{c c c c c}  
\toprule
{Instrument} & {Sample freq. [Hz]} & {Moving avg. [N]} & {Synchronization} & {Common freq. [Hz]} \\[0.5ex]
\midrule
Range meter & 125 & 200 & $\blacklozenge$ & 1000 \\[0.5ex]
Load cell & 5000 & 500 & $\blacklozenge$ & 1000 \\[0.5ex]
Accelerometer & 5000 & 500 & $\blacklozenge$ & 1000 \\[0.5ex]
IMU & 10 & - & $\blacklozenge$ & - \\[0.5ex]
ADCP & 8 & 10 & $\bigstar$ & 80 \\[0.5ex]
ADV & 10 & 10 & $\bigstar$ & 80 \\[0.5ex]
ROV & 30 & - & - & - \\[0.5ex]
\bottomrule
\
\end{tabular}
\caption{Instrument configurations and synchronization. The range meter, IMU, ADCP and ROV were synchronized with Internet time each day. The symbols indicate the instruments that were synchronized in time in the post-processing. } 
\label{Table:instruments} 
\end{table} 

Ice floe motion was also measured with a VN-100 IMU (inertial motion unit) manufactured by VectorNav. The instrument was installed in a rugged box with batteries and a processing unit, see \cite{rabault2020open} for details. The IMU contained a three-axis accelerometer, gyroscope and magnetometer, and allowed for surveillance of all six rigid body motion modes. An integrated GPS tracker provided correct GPS timestamps to the measurements. The sampling frequency was 10~Hz. By examination of the IMU data, it was found that surge was the predominant rigid body motion mode of the ice floe. This is not surprising since the towing was performed in this direction. Some motion was also observed in the other horizontal modes, sway and yaw, i.e. translation in the $y$-direction and rotation about the $z$-axis, respectively, as the floe did not move perfectly parallel to the pool walls. The motion in the vertical modes, heave, roll and pitch, was found to be negligible in comparison with the horizontal motion, and is therefore not further addressed.        

A five beam Nortek Signature1000 (kHz) broadband ADCP was utilized to measure the water velocity in the vicinity of the ice floe. The instrument was operated in the pulse coherent mode, also known as the \textit{high-resolution mode} that enables very small cell size on all beams, which is desirable for turbulence measurements. It was mounted downward facing through a hole in the fast ice from a specially constructed frame, so that the transducer head was 3~cm below the bottom of the ice (i.e. at $z=-3$~cm). The $x$-position was either -0.50~m or -0.25~m and the $y$-position was -0.50~m. In one experiment, the ADCP was placed on the ice floe center. The instrument has one vertically oriented beam $\mathbf{b_5}$, which was pointing in the $-z$-direction, and four slanted beams $\mathbf{b_1-b_4}$ diverging at $\theta=25^{\circ}$ from the vertical. The horizontal components of $\mathbf{b_1-b_4}$ were pointing in the $x$, $y$, $-x$ and $-y$- direction, respectively, as seen in figure~\ref{fig:setup_sketch}. Water velocity along the five beam directions (positive direction was radially away from the instrument) is denoted $b_j$ for $j=1,2,...,5$. 

The mean horizontal velocity components due to the tidal current (measured when the floe was not moving) $\langle u \rangle$ and $\langle v \rangle$, corresponding to $x$ and $y$-directions, respectively, were calculated as $\langle u \rangle = \langle b_1 \sin(\theta) - b_3 \sin(\theta) \rangle $ and $\langle v \rangle = \langle b_2 \sin(\theta) - b_4 \sin(\theta) \rangle$, where the angle brackets denote time averaging over the duration of the time series. The mean horizontal current speed $U_{mean}$ was calculated as $U_{mean} = \sqrt{\langle u \rangle^2+\langle v \rangle^2}$. The ADCP measurement rate was 8~Hz, which is the maximum possible sampling frequency when all the beams are operated. A blanking distance of 10~cm was applied to avoid transducer ringing. The profiling range was 1.9~m and the bin size was either 2 or 5~cm, which yielded 95 or 39 bins, respectively. The instrument settings and placement are summarized in table~\ref{Table:runs}.  



In order to validate the data from the current profiler, a 5~MHz SonTek Hydra ADV was deployed next to the ADCP. The instrument was mounted through a second hole in the fast ice with the measurement volume centered 58~cm below the bottom of the ice. The two acoustic instruments were situated in the same $x$-position and the $y$-position of the ADV was carefully selected so that its measurement volume was very close to the ADCP $\mathbf{b_4}$, as illustrated in figure~\ref{fig:setup_sketch}. The short distance between the two instruments was possible due to the different acoustic frequencies. The ADV was configured with a fixed measurement interval with 10 min continuous sampling followed by 1.67 min down-time. Consequently, not all the cycles were sampled if the instrument down-time coincided with the experiment. Table~\ref{Table:runs} lists the total amount of cycles and cycles sampled by the ADV in the experiments. The ADV measurement frequency was 10~Hz. It was configured to output $ENU$ (east, north and up) velocity components, which were converted to $u$ and $v$-components corresponding to $x$ and $y$-directions, respectively, according to $u = N \cos(\alpha) - E \sin(\alpha)$ and $v = -N \sin(\alpha) - E \cos(\alpha)$. The $w$-component corresponding to the $z$-direction was simply $w = U$. The ADV velocity component $vw$ corresponding to the ADCP $b_4$ velocity was calculated as $vw=-v \sin(\theta) - w \cos(\theta)$, which enabled a direct comparison of the time series from the two instruments on the location indicated by the red square in figure~\ref{fig:setup_sketch}. 



\begin{figure}
  \begin{center}
    \includegraphics[width=.98\textwidth]{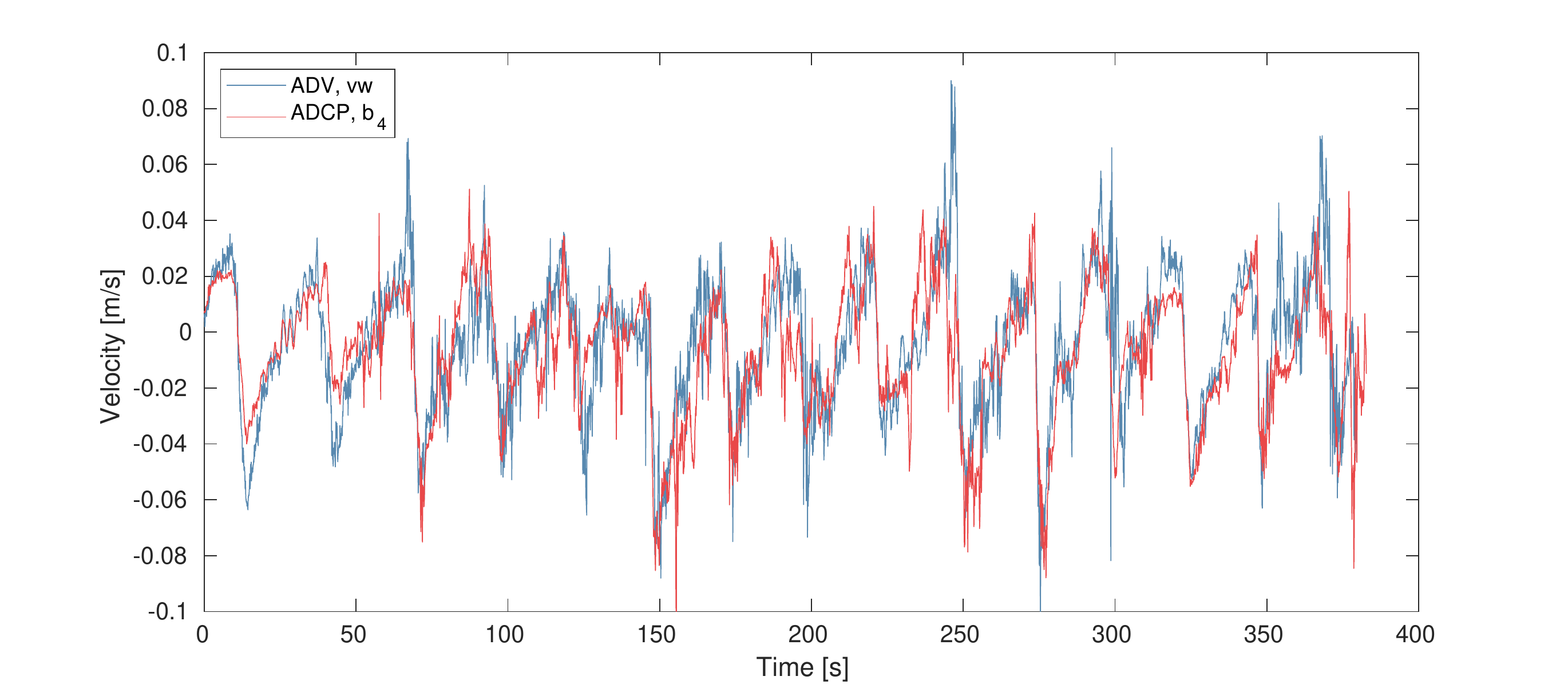} 
    \caption{\label{fig:timeseries_AD} Time series from acoustic Doppler instruments in Exp.~1. The ADCP $b_4$ is from the bin which is located at the same $z$-position as the ADV measurement volume. The ADV $vw$ is the velocity component corresponding to the along-beam velocity of the ADCP $\mathbf{b_4}$. The two measurement volumes were placed as close to each other as possible. }
  \end{center}
\end{figure}

In the post-processing, the ADCP and ADV data were re-sampled to a common sampling rate of 80~Hz and synchronized in time with a cross-correlation optimization method (marked with stars in table~\ref{Table:instruments}), see \cite{loken2021investigation} for further details. An example of a time series from Exp.~1 is shown in figure~\ref{fig:timeseries_AD}, where the ADCP $b_4$ from the bin closest to the ADV measurement volume and the ADV $vw$ are presented. The instruments agree quite well on the larger turbulent scales, but there are some discrepancies, especially on the smaller scales. The two presented time series were recorded spatially very close to each other, but there is of course a limit to how accurately instruments can be placed in field experiments, and there may have been small variations in the ice thickness which led to small errors in the estimated position. In addition, the measurement volumes are different for the ADV and the ADCP, in the order of 1 and 100~$\mathrm{cm^3}$, respectively. The large-scale fluctuations indicate that the ice floe undergoes 15 full cycles. 


In addition to acoustic measurements, fluid motion was also visualized with bubbles as tracing particles. Bubbles were generated from a thin, 0.5~m long carbon fiber pipe perforated every 1~cm on the upward facing side with a 0.1~mm drill bit. The pipe was fed with air of approximately 0.4~bar from a compressor, via a 5~m long flexible rubber hose. This configuration provided an array of bubbles with approximately 2-3~mm diameter. The bubble pipe was attached to the bottom of a metal grid, which was suspended below the ice from strings of thin rope. The bubble pipe was hanging horizontally, aligned with the $x$-axis at $z\approx -1$~m. Bubble motion was recorded with the camera of a BlueROV2 \citep{BlueRobotics} remotely operated vehicle, which was steered below the ice with the camera axis perpendicular to the bubble plane. The frame rate was 30~frames/s and other camera settings such as exposure, brightness and gain were adjusted to ensure that the bubbles appeared as clear, circular particles. The setup, which is illustrated in figure~\ref{fig:setup_sketch}, is further described and validated in \cite{loken2021bringing}.



\subsection{Turbulence analysis} \label{subsec:Turbulence}                

Beam correlation is a quality indicator for acoustic velocimeters, which should exceed 50\% for the ADCP and 70\% for the ADV per manufacturer recommendation. Some spikes occurred in the time series, typically where the correlation dropped below the recommended values. Spikes were identified as velocities outside a range of the moving mean velocity, which was calculated over a sliding window of 10 data points \citep{marchenko2021field}, $\pm$~3 times the standard deviation \citep{nystrom2007evaluation}. For the spectral analysis, which requires continuous time series, the identified spikes were cut where they exceeded the moving mean velocity $\pm$~3 times the standard deviation. In calculations of statistical parameters, such as variance, the spikes were discarded. The fluctuating velocity component in any direction $u'_i=u_i- \langle u_i \rangle$, where $\langle u_i \rangle$ is the time average over the whole time series, was used in the turbulence analysis. For the comparison of turbulent properties obtained from the ADCP and the ADV, time series containing the same number of cycles were used in the analysis, even though the ADCP sampled all the cycles in the experiments (see table~\ref{Table:runs}). 

Turbulent kinetic energy frequency spectra, also known as power spectral densities $PSD_{w}(f)$, where $f$ is the frequency, were estimated from the vertical fluctuating velocity component $w'$ with the Welch method \citep{earle1996nondirectional}, which means fast Fourier transformation and ensemble averaging of overlapping segments. Each time series was divided into 50~s segments with 50\% overlap and a Hamming window was applied to each segment to reduce spectral leakage. Depending on the number of cycles recorded in each experiment (5-20), the resulting spectra had approximately 6-28 degrees of freedom. The TKE frequency spectra represent the distribution of turbulent kinetic energy over the frequencies $0<f<f_N$, where $f_N$ is the Nyquist frequency, which was 4 and 5~Hz for the ADCP and the ADV, respectively. 

In a general perspective of solid-fluid interactions, energy is transferred from the shear flow to large turbulent structures, i.e. the low frequency eddies, where TKE is produced. The low frequency turbulence is considered anisotropic due to the flow geometry, but as energy cascades to the increasingly smaller structures, the directional dependence is lost, and the turbulence is considered locally isotropic and homogeneous. Energy is eventually transfered to the smallest structure of the flow where it is dissipated into heat due to viscosity. The velocity measurement in frequency is related to the turbulent wavenumber $k$ through the velocity $\langle w_{adv} \rangle=2\pi f/k$, that is the time averaged vertical speed at which the turbulence advect past the measurement instrument. Due to the cyclic flow in the present experiment, $\langle w_{adv} \rangle$ was nearly zero and is therefore substituted with $w_{rms}$, which is the root mean square value of the fluctuating vertical velocity component \citep{tennekes1975eulerian,zippel2018turbulence}. In the inertial subrange, the flow is assumed locally isotropic and the TKE frequency spectra should be proportional to $f^{-5/3}$ according to the Kolmogorov law for developed turbulence \citep{kolmogorov1941the}. Within the inertial subrange, the TKE frequency spectra depend only on the TKE dissipation rate $\epsilon$ and $f$, which represents the structure size

\begin{equation}
\label{eq:epsilon}
PSD_{w}(f) = C_K \epsilon^{2/3} f^{-5/3} \left( \frac{w_{rms}}{2\pi} \right)^{2/3},
\end{equation}
             
\noindent where $C_K$~=~0.53 is the universal Kolmogorov constant \citep{sreenivasan1995universality}. Equation~\ref{eq:epsilon} implies that $\epsilon$ can be estimated from the TKE spectra \citep{lumley1983kinematics}, provided that the inertial subrange is resolved by the instruments. 

For the ADCP, a spectrum was estimated for each bin along the vertical beam. Following \cite{guerra2017turbulence}, $\epsilon$ was estimated by solving $\overline{PSD_{w}(f)f^{5/3}|_{f_1}^{f_2}}=C_K \epsilon^{2/3} (w_{rms}/2\pi)^{2/3}$, where $f_1$~=~0.2 to $f_2$~=~1.0~Hz is the range of frequencies with a slope close to zero in the compensated spectrum $PSD_{w}(f)f^{5/3}$, which should be flat in the inertial subrange, and the horizontal bar denotes averaging over the range of frequencies between $f_1$ and $f_2$ (indicated by the vertical bar). The uncertainty in the estimated TKE dissipation rate $\sigma_\epsilon$ is expressed by propagating the uncertainty in the compensated spectrum     

\begin{equation}
\label{eq:epsilon_sigma}
\sigma_\epsilon = \frac{3\pi}{w_{rms}C_K^{3/2}} \sigma_{comp} \sqrt{\overline{PSD_{w}(f)f^{5/3}|_{f_1}^{f_2}}} ,
\end{equation}
             
\noindent where $\sigma_{comp}$ is the standard deviation of the compensated spectrum over the range of frequencies $f_1-f_2$ \citep{guerra2017turbulence}. 

Acoustic instruments have intrinsic Doppler noise $n$ in the beam velocity measurements, which is caused when the Doppler shift is estimated from finite-length pulses \citep{voulgaris1998evaluation}. The Doppler noise often results in flat TKE frequency spectra, also known as the noise floor, typically towards the higher frequencies where the turbulent energy is low. From inspections of both ADCP and ADV data, it was observed that the noise floor was reached close to the Nyquist frequency. Therefore, the noise floor was found by averaging the 20 highest frequencies of the TKE spectra, which corresponds to frequencies in the range 3.7-4 and 4.6-5~Hz for the ADCP and the ADV, respectively. Following \cite{thomson2012measurements}, the noise variance $n^2$ was estimated by integrating the noise floor over the range of frequencies $0<f<f_N$, assuming white noise spectra. The Doppler noise can vary with flow speed and distance from the transducer, so the ADCP noise variance was therefore estimated for all beams and bins.     

The velocity variance $\langle u'^{2}_{i} \rangle$ was obtained by squaring and time averaging the fluctuating velocity components. The Doppler noise was removed from the velocity variance statistically \citep{lu1999using} by subtracting the noise variance, so that $\langle u'^{2}_{i} \rangle=var(u'_i)-n^2$. Instances that were considered to be spikes or with correlation less than the recommended values were removed from the time series before the calculations of the velocity variance were made. Following \cite{dewey2007reynolds}, the total TKE density $TK$ was calculated as

\begin{equation}
\label{eq:TKE_ADV}
TK_{ADV} = \rho_w\frac{\langle u'^2 \rangle + \langle v'^2 \rangle + \langle w'^2 \rangle}{2},
\end{equation}

\begin{equation}
\label{eq:TKE_ADCP}
TK_{ADCP} = \rho_w\frac{\langle b'^{2}_{1}\rangle + \langle b'^{2}_{2}\rangle + \langle b'^{2}_{3}\rangle + \langle b'^{2}_{4}\rangle - 2(2 \cos^2 \theta - \sin^2 \theta) \langle b'^{2}_{5}\rangle}{4 \sin^2 \theta},
\end{equation}

\noindent for the ADV and the ADCP, respectively.


\section{Theoretical background} \label{sec:theoretical_background}

\begin{figure}
  \begin{center}
    \includegraphics[width=.65\textwidth]{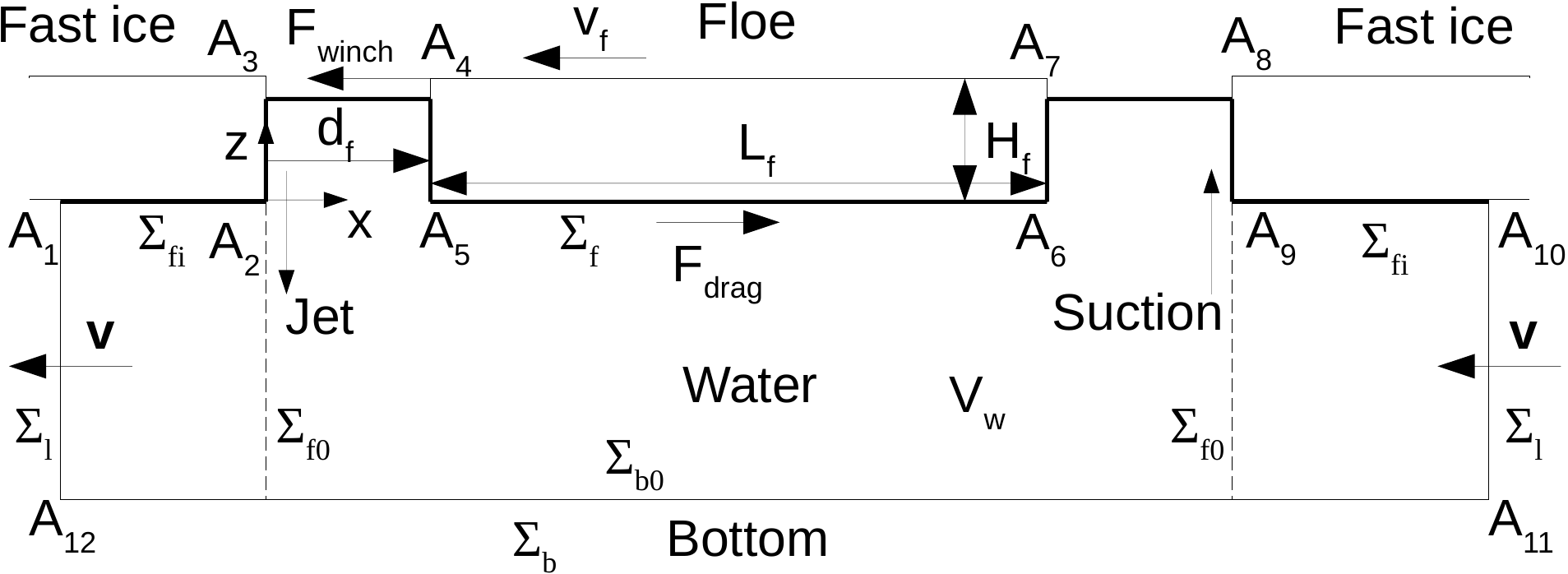}
    \caption{\label{fig:momentum} Scheme of the fast ice and floe and the liquid domain below.}
  \end{center}
\end{figure}

In this section, the moving ice floe and the surrounding water are described theoretically, and the different forces acting on the floe and the mechanisms of energy input and dissipation are identified. An idealized sketch of the towing situation is presented in figure~\ref{fig:momentum}, where an ice floe is free to move inside a pool in the fast ice. The towing force applied by the winch $F_{winch}$ initiate floe motion and act in the same direction as the axial floe velocity $v_{f,x}$ (at the gravity center of the floe), whereas the frictional forces applied on the ice floe by the surrounding water $F_{drag}$ act in the opposite direction. Similarly, power $P$ is transferred to the floe from the winch ($P_{winch}$) and to the water from the floe ($P_{drag}$) due to the external forces $F$, where $P = Fv_{f,x}$. The energy balance of the floe can be described by


\begin{equation}
\label{eq:E_balance_floe}
\frac{dK_f}{dt} = P_{winch} - P_{drag} - P_{coll} - P_{other},
\end{equation}

\noindent where $K_f = \sum_{i=1}^{3}(mv_{f,i}^2+I_i \omega_{f,i}^2)/2$ is the kinetic energy of the floe, $m$ and $I$ are the mass and moment of inertia of the floe, respectively, $\omega_f$ is the angular velocity of the floe rotation around the gravity center, $t$ is time, $P_{coll}$ is the power dissipated in the floe collisions with the fast ice and $P_{other}$ is the power dissipated in other processes, such as losses in the towline and ice screws. Equation~\ref{eq:E_balance_floe} is equal to zero when it is time averaged over the period of the oscillating motion.

Now, the water volume $V_w$ around and below the fast ice and floe bounded by the broken line $A_1-A_{12}$ in figure~\ref{fig:momentum} is considered. The volume boundary $\Sigma$ consists of the boundary with the fast ice $\Sigma_{fi}$, the boundaries of the pool with the floe passing the points $A_2-A_9$, the lateral boundaries of the water volume $\Sigma_l$ and the bottom boundary associated with the seabed $\Sigma_b$. The submerged surface of the floe $\Sigma_f$ consists of the broken line $A_4-A_7$. The sea depth is constant, and the fast ice is extended horizontally to the infinity from the pool. It is assumed that a large-scale pressure gradient associated with the semi-diurnal tide influences the sea current below the ice with a mean horizontal velocity $\mathbf{v} = \mathbf{v}(z)$, which generates the background turbulence. According to \cite{landau2013course} (Eq.~16.1), the kinetic energy balance of the water inside the volume $V_w$ is written as follows 

\begin{equation}
\label{eq:E_balance_water}
\frac{dK_w}{dt} = \int_{\Sigma} [\boldsymbol{\sigma}_n - K\mathbf{n}] \cdot \mathbf{v} dS - D_v,
\end{equation}  

\noindent where $K_w$ is the kinetic energy of the water, $\boldsymbol{\sigma}_n$ is the stress vector and $\mathbf{n}$ is the outward unit normal vector at the boundary $\Sigma$, $K$ is the density of kinetic energy, $\mathbf{v} = (u,v,w)$ is the water velocity and $D_v$ is the rate of viscous energy dissipation. The kinetic energy of the water is determined as $K_w = \int_{V_w} K dV$, where $K = \rho_w(u^2+v^2+w^2)/2$ and $\rho_w$ is the water density. The rate of viscous energy dissipation is determined by the formula $D_v = \mu \int_{V_w} (\partial v_i/\partial x_j + \partial v_j/\partial x_i)^2 dV/2$, where $\mu$ is the dynamic viscosity of water.     

First, the case of a mean steady flow due to the tidal current below a continuous fast ice is considered, where $dK_w/dt = 0$ and $K=const$. In this situation, i.e. in the absence of an ice floe, Eq.~\ref{eq:E_balance_water} leads to

\begin{equation}
\label{eq:E_balance_water_cont}
\int_{\Sigma_l} p\mathbf{n} \cdot \mathbf{v} dS + D_v = 0,
\end{equation}  

\noindent where $p$ is the water pressure. Equation~\ref{eq:E_balance_water_cont} states that the work of water pressure equals the energy dissipation inside the water volume $V_w$. The integral in Eq.~\ref{eq:E_balance_water_cont} is negative because the water moves in the opposite direction to the pressure gradient. The remaining terms from Eq.~\ref{eq:E_balance_water} are zero. It is assumed that $\boldsymbol{\sigma}_n=0$ at the open surface of water between the floe and the fast ice. The integral of $K\mathbf{n} \cdot \mathbf{v}$ equals zero if the surface $\Sigma_l$ is extended far away from the pool where the influence of the floe on the sea current is small: the integral of $K\mathbf{n} \cdot \mathbf{v}$ equals zero due to symmetry over $\Sigma_l$ and because $\mathbf{n} \cdot \mathbf{v}=0$ at the ice, water and bottom surface over $\Sigma_{fi}$ and $\Sigma_{b}$.      

Next, the ice floe with the periodic back and forth motion is introduced. In this section, angled brackets are used to describe time averaging over the period of the oscillating motion. Equation~\ref{eq:E_balance_water} is averaged over the period of the oscillating motion, which leads to

\begin{equation}
\label{eq:E_balance_water_floe}
\int_{\Sigma_l} \langle p\mathbf{n} \cdot \mathbf{v} \rangle dS - \int_{\Sigma_f} \langle \boldsymbol{\sigma}_n \cdot \mathbf{v} \rangle dS + \langle D_v \rangle = 0,
\end{equation}  

\noindent where the second integral is equal to the power of the floe work to move the surrounding water $P_{drag}=\int_{\Sigma_f} \langle \boldsymbol{\sigma}_n \cdot \mathbf{v} \rangle dS$, and $\langle D_v \rangle$ is the average rate of energy dissipation. 

It is assumed that $\langle p\mathbf{n} \cdot \mathbf{v} \rangle \approx p\mathbf{n} \cdot \mathbf{v}$ over the lateral surface $\Sigma_l$ in Eqs.~\ref{eq:E_balance_water_cont}-\ref{eq:E_balance_water_floe} if $\Sigma_l$ is extended far away from the pool where the influence of the floe motion is small. Subtraction of Eq.~\ref{eq:E_balance_water_cont} from Eq.~\ref{eq:E_balance_water_floe} leads to the equation

\begin{equation}
\label{eq:P_drag_water}
P_{drag} = \int_{\Sigma_{f}} \boldsymbol{\sigma}_n \cdot \mathbf{v} dS = \langle D_v \rangle - D_v.
\end{equation}  

The dissipation rates $\langle D_v \rangle$ and $D_v$ can be written as integrals $\langle D_v \rangle = \int_{\Sigma_b} \langle d_v \rangle dx dy$ and $D_v = \int_{\Sigma_b} d_v dx dy$, where $\langle d_v \rangle$ and $d_v$ are the area densities of the energy dissipation rates, and $x$ and $y$ are the horizontal coordinates. It is assumed that $\langle d_v \rangle = d_v$, far away from the floe. The difference $\langle D_v \rangle - D_v$ can be written as a sum

\begin{equation}
\label{eq:D_water}
\langle D_v \rangle - D_v = \int_{\Sigma_{b0}} (\langle d_v \rangle - d_v) dx dy + \int_{\Sigma_{l0}} (\langle K\mathbf{n} \cdot \mathbf{v} \rangle - K\mathbf{n} \cdot \mathbf{v}) dS,
\end{equation} 

\noindent where $\Sigma_{b0}$ is the part of the sea bottom surface below the pool and $\Sigma_{l0}$ is the vertical cylindrical surface separating the pool from the fast ice. The first integral on the R.H.S of Eq.~\ref{eq:D_water} describes the energy dissipation rate in the water bounded by the surface $\Sigma_{l0}$, and the second integral equals the kinetic energy transported through the surface $\Sigma_{l0}$ by the sea current in unit time and dissipated outside the surface $\Sigma_{l0}$. 


Equation~\ref{eq:P_drag_water} expresses $P_{drag}$ from the dissipation rate in the water, which is estimated from the ADCP data in Section~\ref{subsec:TKE_energy}, but it may also be described as the power of the drag force $F_{drag}$. The power of the ice floe work to move the surrounding water can be performed conditionally as a sum of the power of normal pressure stresses $P_{fd}$ and tangential shear stresses $P_{sd}$, i.e. due to form drag $F_{fd}$ and skin friction drag $F_{sd}$, respectively (e.g. \cite{newman2018marine}), and is expressed as

\begin{equation}
\label{eq:P_drag_floe}
P_{drag} = P_{fd} + P_{sd} = - \int_{\Sigma_{f}} \langle p\mathbf{n} \cdot \mathbf{v} \rangle dS + \int_{\Sigma_{f}} \langle \boldsymbol{\tau}_n \cdot \mathbf{v} \rangle dS.
\end{equation}  

An attempt is made to estimate $P_{drag}$ based on empirical drag coefficients and the axial floe displacement $d_{f,x}$ and velocity $v_{f,x}$ shown in figure~\ref{fig:momentum}. Viscous drag forces are proportional to the squared relative velocity between the floe and the ambient water flow. However, it will be shown in Section~\ref{subsec:TKE_energy} that the ambient water velocity was small compared to the floe velocity, and its predominant direction was perpendicular on the towing direction, hence negligible in the context of drag forces. Therefore, drag forces are considered to be proportional to the squared floe velocity. The power due to skin friction drag $P_{sd}$ is estimated as 

\begin{equation}
\label{eq:P_skin}
P_{sd} \approx \rho_w S_{sd} C_{sd} v_{f,x}^3,
\end{equation} 

\noindent where $\rho_w=1026~\mathrm{kg}\mathrm{m}^{-3}$, $S_{sd} = L_f W_f$ is the area of the floe bottom and $C_{sd}$ is the skin friction drag coefficient. Ideally, the surface of the lateral floe sides $2L_fH_f$ should be included in $S_{sd}$ since a skin drag force will be applied here as well. However, the water velocity was not measured in the gap between the lateral floe sides and the fast ice, thus this contribution to $P_{sd}$ is neglected in the calculations for the comparison with the measurements. The skin friction drag coefficient $C_{sd}$ is related to the ratio $L_f/H_f$ and is set to $1.5\times 10^{-2}$ \citep{bai2006optimization}. Form drag is estimated as $F_{fd} \approx \rho_w S_{fd} C_{fd} v_{f,x}^2/2$, where $S_{fd} = W_f H_f$ is the area of the floe vertical cross section with normal vector parallel to the axial direction and $C_{fd}$ is the form drag coefficient, which is set to 1 \citep{hoerner1965fluid}.   

Consider one half towing cycle, where the floe starts by one of the pool walls $A_8-A_9$ at $t=t_0$ and collides with the opposite pool wall $A_2-A_3$ at $t=t_2$. The empirical formula on form drag stated above may be insufficient when the floe approaches the fast ice immediately before $t_2$ due to the increasing water pressure in the decreasing volume $V_{w,j} = d_{f,x}W_fH_f$ (bounded by the broken line $A_2-A_5$). The floe transfers kinetic energy to the surrounding water when the water in $V_{w,j}$ exits the surface $S_{j}=d_{f,x}(W_f+2H_f)$ as a jet. The power of normal pressure stresses $P_{fd}$ is assumed equal to the jet power $P_{jet}$ in the time span between $t_1$ and $t_2$ immediately before collision. From the ROV images, the jet appears when $d_{f,x} \ll H_f$, so $t_1$ is set to be the instance when $d_{f,x}(t_1) = 0.1~\mathrm{m}$. The jet power is estimated as $P_{jet} \approx \rho_w (W_fH_f v_{f,x})^3/2S_j^2$.


It is assumed that the jet power substitutes the form drag power in the time span between $t_1$ and $t_2$, so that

\begin{equation}
\label{eq:P_form}
P_{fd} \approx \frac{\int_{t_0}^{t_1} v_{f,x}F_{fd} dt + 2E_{jet}}{t_2-t_0},
\end{equation}

\noindent where $E_{jet} = \int_{t_1}^{t_2} P_{jet} dt$ is the jet kinetic energy. On the opposite side of the floe, a suction of water will be induced into the opening gap immediately after $t_0$. Due to symmetry arguments, it is assumed that the energy transfer to the water through the suction motion $E_{suction}$ equals $E_{jet}$, hence is $E_{jet}$ in Eq.~\ref{eq:P_form} multiplied by 2.     

Range meter data from Exp.~3 is applied to calculate the energy dissipated by the frictional forces. Ice floe displacement $d_{f,x}$ and velocity $v_{f,x}$ are similar to the values presented in the upper panel of figure~\ref{fig:timeseries_evo}, except that $d_{f,x}=0$ is placed 5~cm from the pool wall because visual observations and the ROV images (presented in figure~\ref{fig:plume_evolution}) revealed that ice was forming around the floe and pool circumference at the water surface, which prevented the ice floe surface to approach the fast ice closer than approximately 5~cm. The average power transfer from the floe to the surrounding water $P_{drag} = P_{fd} + P_{sd}$ is estimated to be 11.6~W with Eqs.~\ref{eq:P_skin}-\ref{eq:P_form}. 


\section{Results} \label{sec:Results}

The results are organized according to Eq.~\ref{eq:E_balance_floe} in Section~\ref{sec:theoretical_background}, i.e. as an energy balance of the system of interest, consisting of the ice floe and the surrounding water bounded by the fast ice. The power input to the system from the electrical winches $P_{winch}$ is compared with the rate of energy dissipation in the floe-wall collisions $P_{coll}$ and the total TKE rate in the surrounding water due to the floe motion, which is equivalent to $P_{drag}$. The two former terms are calculated as an average amount of energy, either as input or consumed per half cycle, and divided by the average duration of a half cycle to obtain the unit of power, whereas the latter term is estimated from time series of the entire experiments and is expressed as rate of energy dissipation. The reader is reminded that six experiments are included in this paper, and that each experiment contained around 10 periods of ice floe towing oscillations, referred to as cycles. 


\subsection{Input energy} \label{subsec:Input_energy}

\begin{figure}
  \begin{center}
    \includegraphics[width=.98\textwidth]{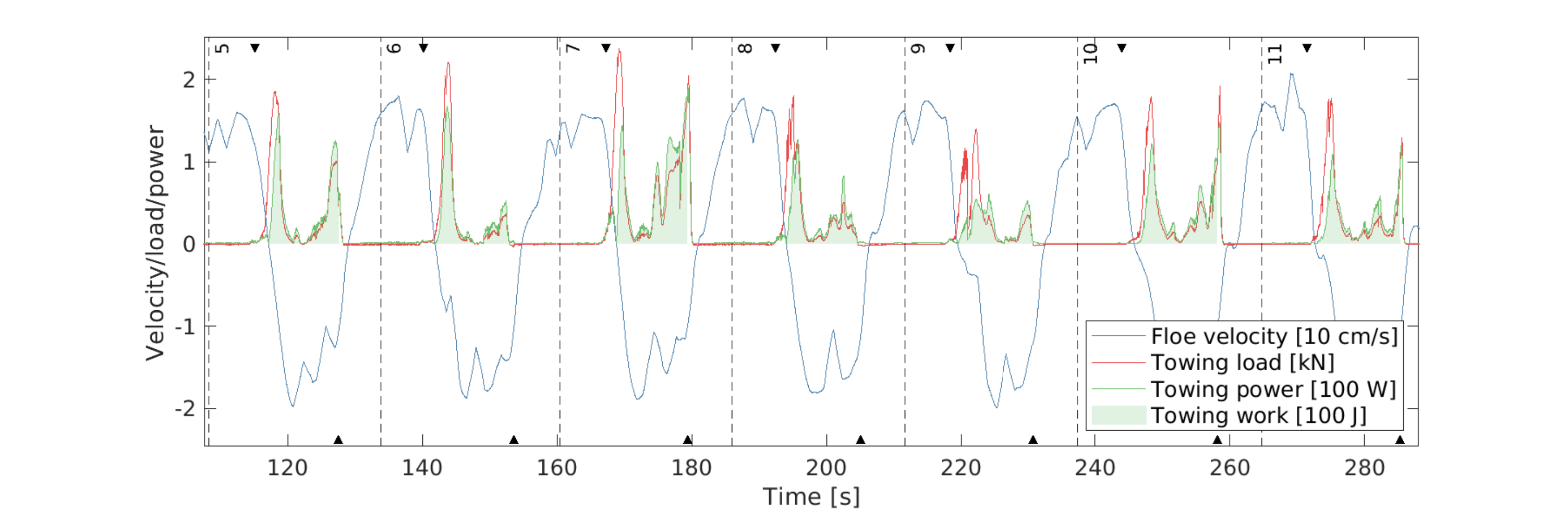}
    \caption{\label{fig:input_work} Part of the time series of ice floe translational velocity in the axial direction (blue), towing load (red) and towing power (green) applied by the south end winch, including Cycle~5-11 in Exp.~3. Shaded regions indicate towing work. Upward and downward-pointing triangles indicate the time of the collisions on the south and north ends, respectively.}
  \end{center}
\end{figure}

Range meter and load cell data were combined to investigate the input energy rate to the system of interest. The instantaneous power input $P_{winch}$ was determined as the product of the floe velocity in the axial direction $v_{f,x}$ and the towing load applied by the winch $F_{winch}$. Figure~\ref{fig:input_work} shows a part of the time series including Cycles~5-11 in Exp.~3 as an example. The cycles are marked with numbers and separated with vertical dashed lines. Negative velocity means displacement towards the south end of the pool. The load cell only provided information when the towing occurred in the $-x$-direction. The work performed by the winch on the ice floe $E_{winch}$ during a half cycle was determined as the integral of the towing power with respect to time over the time span of the half cycle. This corresponds to the shaded areas in figure~\ref{fig:input_work}.  

In each cycle, there was typically one large peak in towing power from accelerating the ice floe, succeeded by a smaller peak. The second peak was probably a consequence of additional power input needed to overcome the increasing water pressure in the closing gap. The shaded areas in figure~\ref{fig:input_work} extend in time until collision occurs. When Cycles~1 and 5-11 are considered (Cycle~2-4 contained severe load cell dropouts), the average work applied to tow the ice floe in one direction $E_{winch}$ was 428 J. One half cycle lasted on average 13.3 s, which means that the average power transfer from the winch to the ice floe was approximately 32.2 W. Due to symmetry arguments, it is assumed that the north end winch applied equal power to the system as the south end winch. The load cell was only applied in Exp.~3. It is assumed that the winch input power to the system was similar in Exp.~1-6 due to the consistency in the towing procedure.

\subsection{Energy dissipation in collisions} \label{subsec:Collision_energy}

\begin{figure}
	\centering
	\subcaptionbox{The south end of the pool.\label{fig:collisions_1}}
		{\includegraphics[width=.49\linewidth]{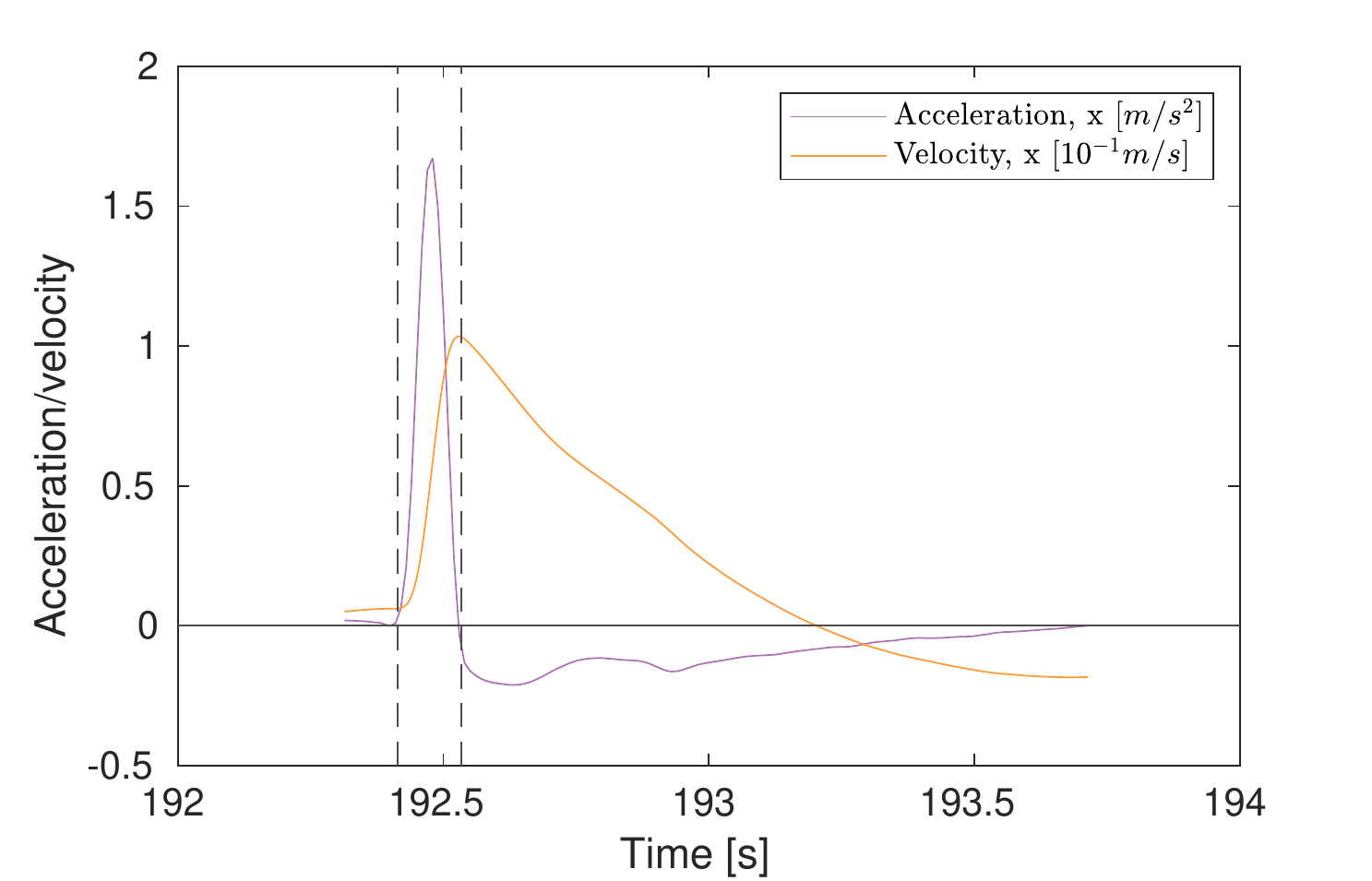}} 
	\subcaptionbox{The north end of the pool.\label{fig:collisions_2}}
		{\includegraphics[width=.49\linewidth]{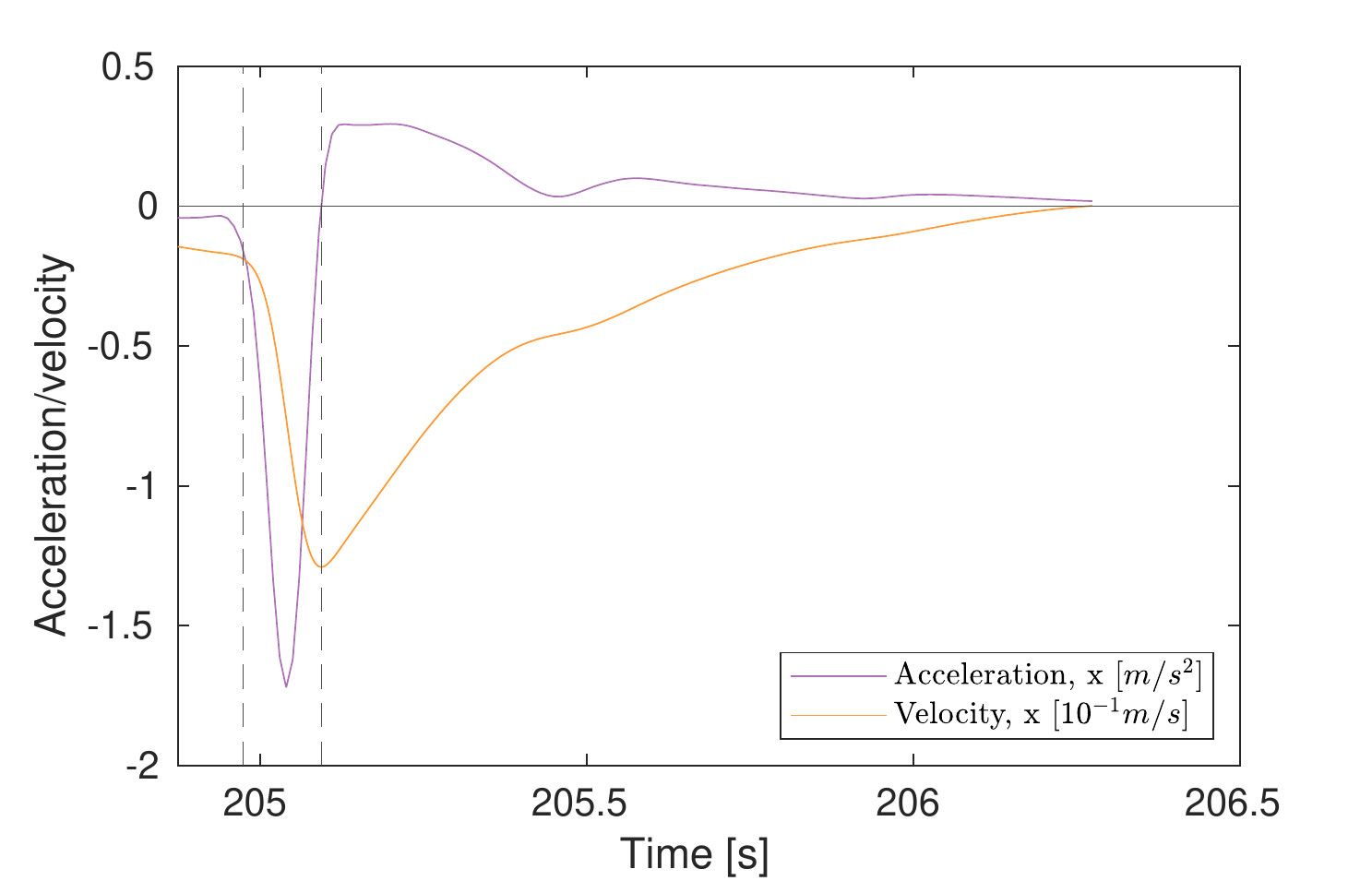}}
	\caption{Collision events during the eighth cycle in Exp.~3. Ice floe acceleration and velocity (integrated acceleration w.r.t. time) in the $x$-direction from the uniaxial accelerometer data. Vertical dashed lines define the start and end of the collisions.}\label{fig:collisions}
\end{figure}

Collisions between the ice floe and the fast ice are characterized from the uniaxial accelerometers placed on the floe. The time series of the acceleration in the $x$-direction from Exp.~3, presented in figure~\ref{fig:timeseries_evo}, reveal periodic recurring spikes, which correspond to impact events. Two events occurred per cycle, when the floe collided in the south and north ends of the pool. Figure~\ref{fig:collisions} presents time series of the acceleration and velocity during the collision events in the eighth cycle of Exp.~3. The velocity was found by numerically integrating the acceleration with respect to time with the cumulative trapezoidal method. After the integration, a second order Butterworth bandpass filter with cutoff frequencies of 0.05 and 100 Hz was applied to remove any low frequency noise associated with the integration \citep{sutherland2016observations}. 


The collision events presented in figure~\ref{fig:collisions} are characterized by an initial peak in the acceleration time series, which corresponds to ice floe deceleration as it approached the ice edge, followed by a smaller acceleration with opposite sign. The latter acceleration may be due to rotation of the floe \citep{marchenko2021field}, which could have happened if the contact faces were not perfectly parallel at the instance of impact $t_{impact}$. Following \cite{li2018laboratory}, the time instance of impact $t_{impact}$ occurs at the peak deceleration, and the collision start and end time, $t_{pre}$ and $t_{post}$, are determined as $t_{impact} \pm \Delta t$, where $\Delta t$ is set to 0.06~s from empirical observations. Hence, the duration of the peak deceleration was 0.12~s ($t_{pre}$ and $t_{post}$ are indicated with vertical dashed lines in figure~\ref{fig:collisions}) and the entire collision event including the initial peak deceleration and the successive acceleration lasted around 1~s. \cite{marchenko2019influence} found from ice block drop experiments that the typical peak deceleration period was 0.1 and 0.01~s for wet and dry collisions, respectively. The peak deceleration amplitude in the current results is 1-2~$\mathrm{m}\mathrm{s}^{-2}$ and the acceleration time series agree in general with the ice floe towing experiments of \cite{marchenko2021field}. 

Accelerometer data from the IMU were investigated for comparison and processed in the same manner as the uniaxial accelerometer data to find velocity time series. The IMU data agree in general with the uniaxial accelerometer data, although the impacts were more poorly resolved due to the much lower sampling frequency. Consequently, the peak deceleration events appeared smaller and lasted longer than the ones obtained from the uniaxial accelerometers. From evaluation of the peak decelerations, $\Delta t$ was set to 0.2~s for the IMU data.    

A sudden change in velocity can be observed during the time of the peak deceleration $\Delta t$ in figure~\ref{fig:collisions}. Following \cite{li2018laboratory}, the energy dissipated in the inelastic collision between floes $E_{coll}$ can be estimated as the difference in kinetic energy $E_{coll} \approx \Delta K_f$ of the floe at $t_{pre}$ and $t_{post}$. As mentioned in Section~\ref{subsec:Instrumentation}, the vertical modes of floe motion were negligible. Therefore, Eq.~1 of \cite{li2018laboratory}, which describes the total kinetic energy of the floe, can be rewritten as 

\begin{equation}
\label{eq:KE}
K_f \approx \frac{1}{2} m v_{f,x}^2 + \frac{1}{2} m v_{f,y}^2 + \frac{1}{2} I_z \omega_{f,z}^2,
\end{equation}

\noindent where $m$ is the ice floe mass, $v_x$ and $v_y$ are the floe translational velocities in the horizontal plane (related to surge and sway), $I_z$ is the moment of inertia about the vertical axis that goes through the floe center of gravity and $\omega_z$ is the rotational velocity about the vertical axis (related to yaw). The ice floe mass was estimated as $m=\rho_{f}L_fW_fH_f$, where $\rho_{f}=910~\mathrm{kg}\mathrm{m}^{-3}$ is a typical sea ice density \citep{timco1996review}. The moment of inertia was estimated as $I_z = m(L_f^2+W_f^2)/12$, i.e. the tabulated value of a rectangular prism, see e.g. \cite{spiegel1999mathematical}. The first two terms on the R.H.S. of Eq.~\ref{eq:KE} were calculated from both uniaxial accelerometer and IMU data, and the two instruments agreed. The last term was only obtained from the IMU data.  

In terms of lost kinetic energy in the collisions, the contribution from surge motion was found to dominate the contributions from sway and yaw by one and two orders of magnitude, respectively. The latter two terms on the R.H.S. of Eq.~\ref{eq:KE} are therefore neglected in the following. From the uniaxial accelerometer data, the dissipated energy in one collision event $E_{coll}$ was found to be 32.5~J on average over the 11 cycles in Exp.~3. Considering the average duration of a half cycle, the mean power dissipated due to collisions $P_{coll}$ was 2.4~W, which corresponds to 7.5\% of the total input energy rate $P_{winch}$. The accelerometers were deployed together with the load cell, i.e. only in Exp.~3. As mentioned earlier, all the experiments were very consistent in terms of ice floe motion. Hence, it is assumed that the rate of energy dissipated in the collisions was similar in Exp.~1-6.  



\subsection{Optical measurements of jet generation} \label{subsec:ROV_jet}  

\begin{figure}
  \begin{center}
    \includegraphics[width=.98\textwidth]{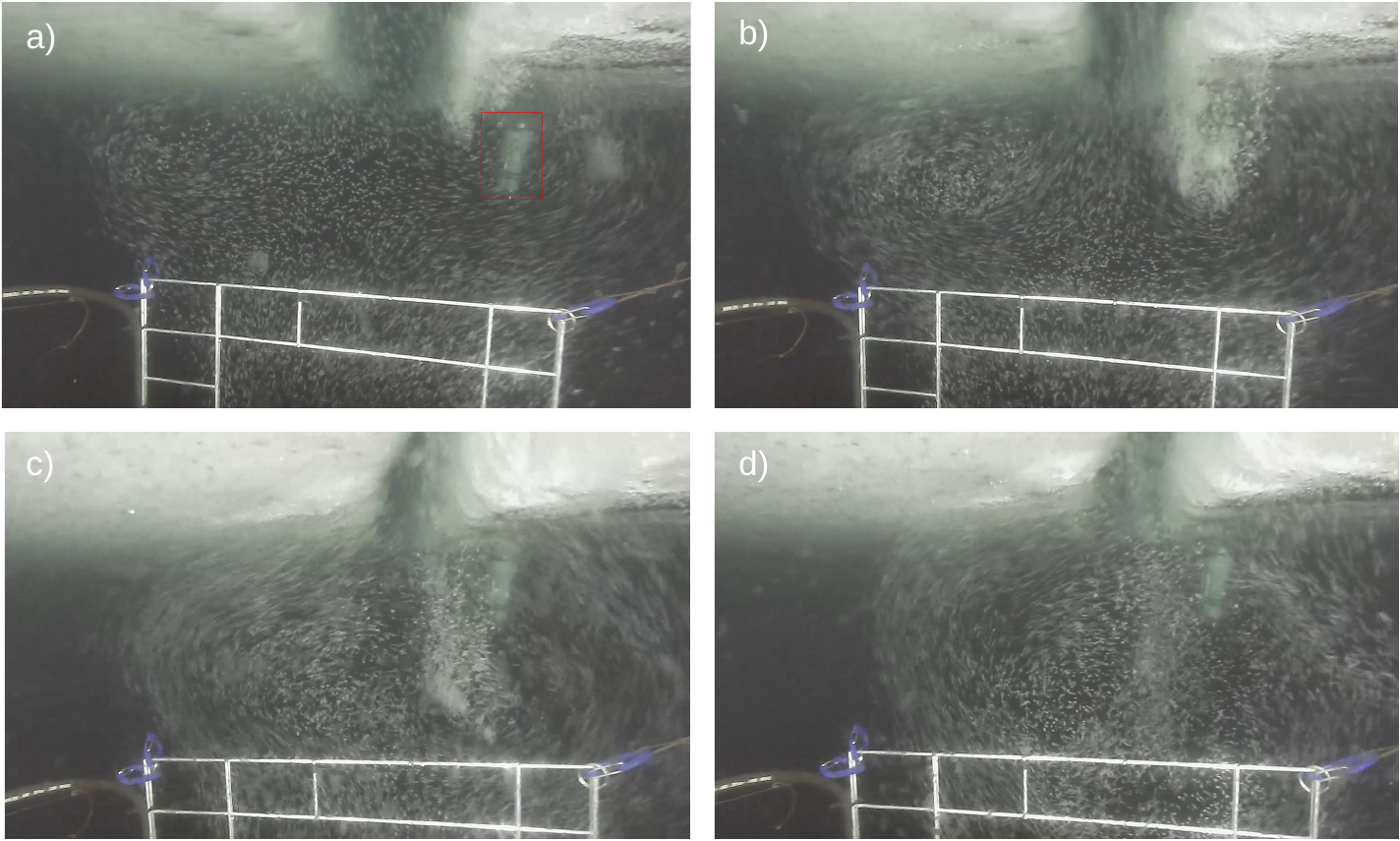}
    \caption{\label{fig:plume_evolution} Evolution of a downward water jet as the ice floe (left) approaches the fast ice (right) during the 11th cycle in Exp.~3. The time span between each frame a)-d) is 0.5~s. The red rectangle indicates the ADV, which was positioned behind the bubble curtain. The ADCP (positioned in front of the ADV) is not visible on the images.}
  \end{center}
\end{figure}

Although the acoustic velocimeters were only deployed on the fast ice next to the pool and in the ice floe center, the ROV and rising bubbles setup provide information on the flow structures below the floe and the fast ice. Figure~\ref{fig:plume_evolution} presents four images taken with the ROV camera, which show the ice floe approaching the south end of the pool during the 11th cycle in Exp.~3. The camera axis is approximately aligned with the $y$-axis. Collision occurs around figure~\ref{fig:plume_evolution}~c. A downward jet is forming in the closing gap with a large eddy structure on each side in the axial direction. The length of the metal grid in the lower part of the image is 0.55~m, meaning that the total jet diameter, including the resulting turbulent cloud, is in the order of 1~m. As the jet evolves, the vortex centers move towards the (vertical) flow axis. The horizontal distance from the flow axis to the vortex center is approximately 0.1-0.3~m.      

\begin{figure}
  \begin{center}
    \includegraphics[width=.75\textwidth]{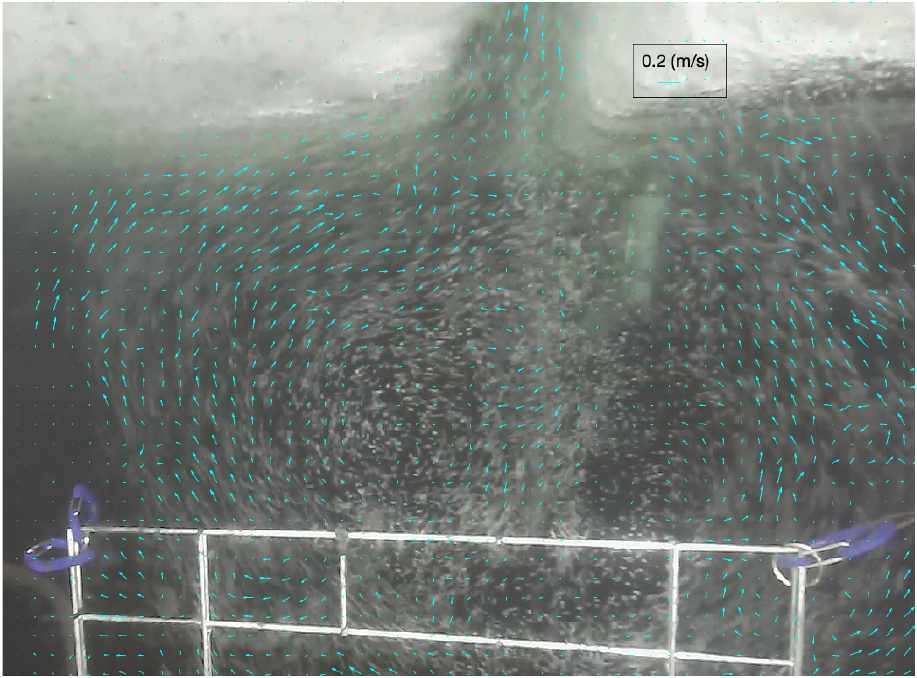}
    \caption{\label{fig:plume_PIV} Downward jet processed with PIV to obtain velocity vectors. The image frame was taken 0.33~s after figure~\ref{fig:plume_evolution}~c. The magnitude of the velocity vectors is indicated in the legend.}
  \end{center}
\end{figure}

Over the last couple of decades, particle image velocimetry (PIV) has been adapted to field experiments to investigate flow kinematics in the ocean, see e.g \cite{smith2002piv,bertuccioli1999submersible,loken2021bringing}. PIV was performed on consecutive ROV image pairs with the in-house HydrolabPIV software developed at the University of Oslo \citep{kolaas2016getting}. The processing was performed with 48$\times$48~pixel subwindows with 50\% overlap. A linear pixel-to-world coordinate transformation was achieved with the mesh-points of the metal grid. The mean vertical buoyancy driven bubble velocity was found in a reference run with calm water and subtracted from the velocity field obtained in the jet. Further details on the experimental setup and processing scheme can be found in \cite{loken2021bringing}. Figure~\ref{fig:plume_PIV} presents the jet 2D velocity field in the $xz$-plane in Exp.~3. As in figure~\ref{fig:plume_evolution}, two large eddies can be seen with centers approximately 10~cm from the flow axis. Circular water motion is evident up to 0.5~m from the flow axis. Smaller turbulent structures were also resolved and can be observed within the jet domain. This observation, particularly the short distance from the flow axis to the vortex center, indicates that the ADCP probably captured the most dominating flow structures when it was placed 0.25~m from the pool edge but may have failed to do so when placed further away.               

\subsection{TKE dissipation} \label{subsec:TKE_energy}

\begin{figure}
  \begin{center}
    \includegraphics[width=.98\textwidth]{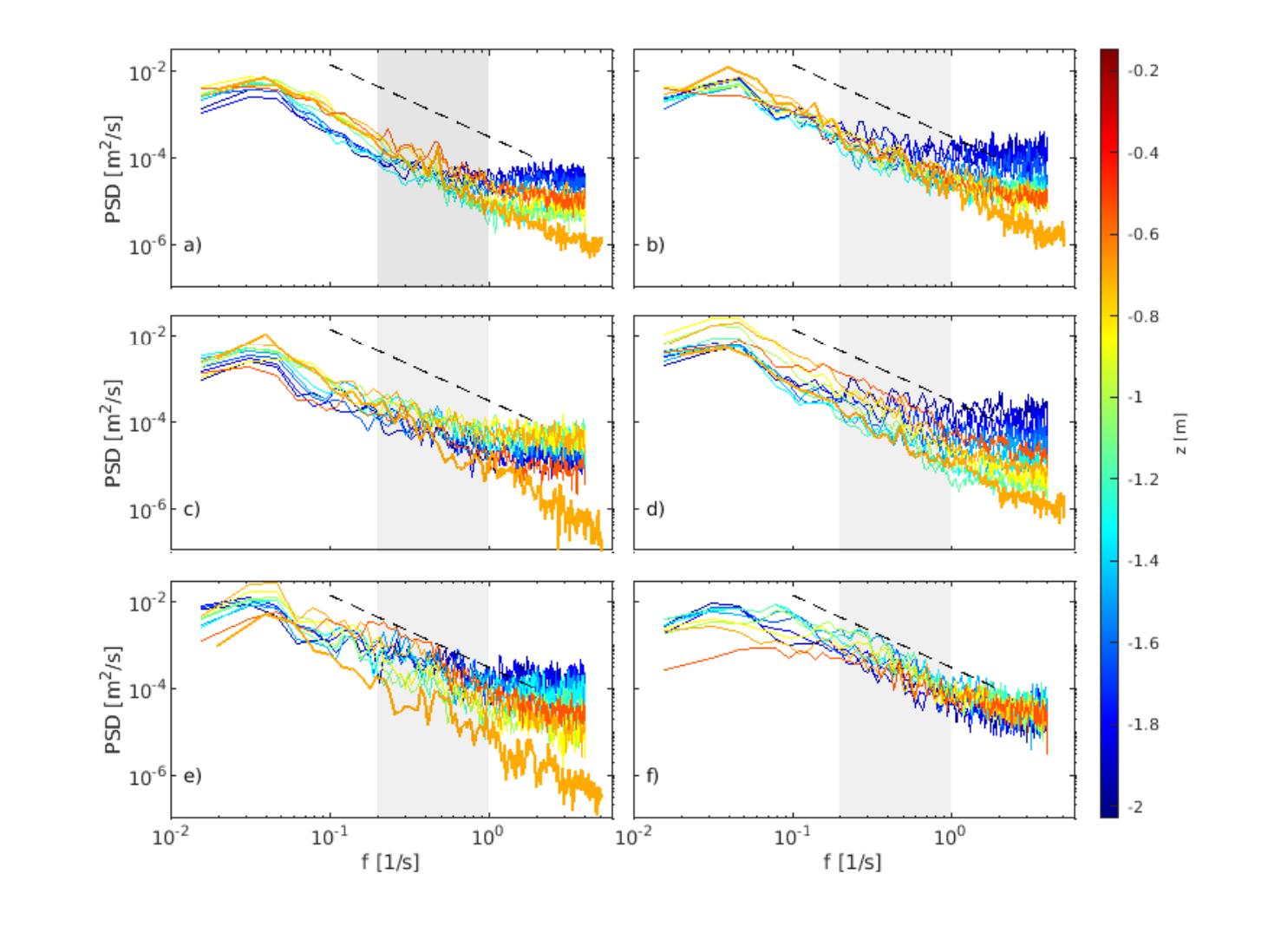}
    \caption{\label{fig:spectra_profile} TKE spectra obtained with the ADCP from various depths in Experiments~1-6 shown in a)-f), respectively. The black dashed lines show the theoretical $f^{-5/3}$ slope. The ADV spectra are shown as thick orange lines in a)-e), but the ADV was not deployed in Exp.~6 when the ADCP was placed in the ice floe center. The shaded regions show the range of frequencies over which the compensated spectra were averaged to estimate $\epsilon$.}
  \end{center}
\end{figure}

From the fluctuating vertical velocity component of the ADV and all the ADCP bins, TKE spectra were estimated with the Welch method described in Section~\ref{subsec:Turbulence}. All the cycles that were measured by both instruments were included in the calculation. The bins corresponding to the 15~cm closest to the instrument head showed some unphysical behavior, probably due to transducer ringing \citep{nystrom2007evaluation}, and were therefore discarded. Figures~\ref{fig:spectra_profile}~a-f present the spectra from Exp.~1-6, respectively, where only 10 ADCP bins evenly distributed over the 2~m deep profile are presented to increase the readability. The thicker orange spectra in Figs.~\ref{fig:spectra_profile}a-e are produced from the ADV, which was not deployed in Exp.~6 when the ADCP was placed in the ice floe center. Most of the spectra exhibit peak frequencies around 0.04~Hz, which correspond to the ice floe surge period of approximately 26~s. The gray shaded regions illustrate the range of frequencies $f_1-f_2$ over which the compensated spectra were averaged in order to estimate the TKE dissipation rate, i.e. where a slope proportional to $f^{-5/3}$ is expected in accordance with Eq.~\ref{eq:epsilon}. 

\begin{figure}
  \begin{center}
    \includegraphics[width=.9\textwidth]{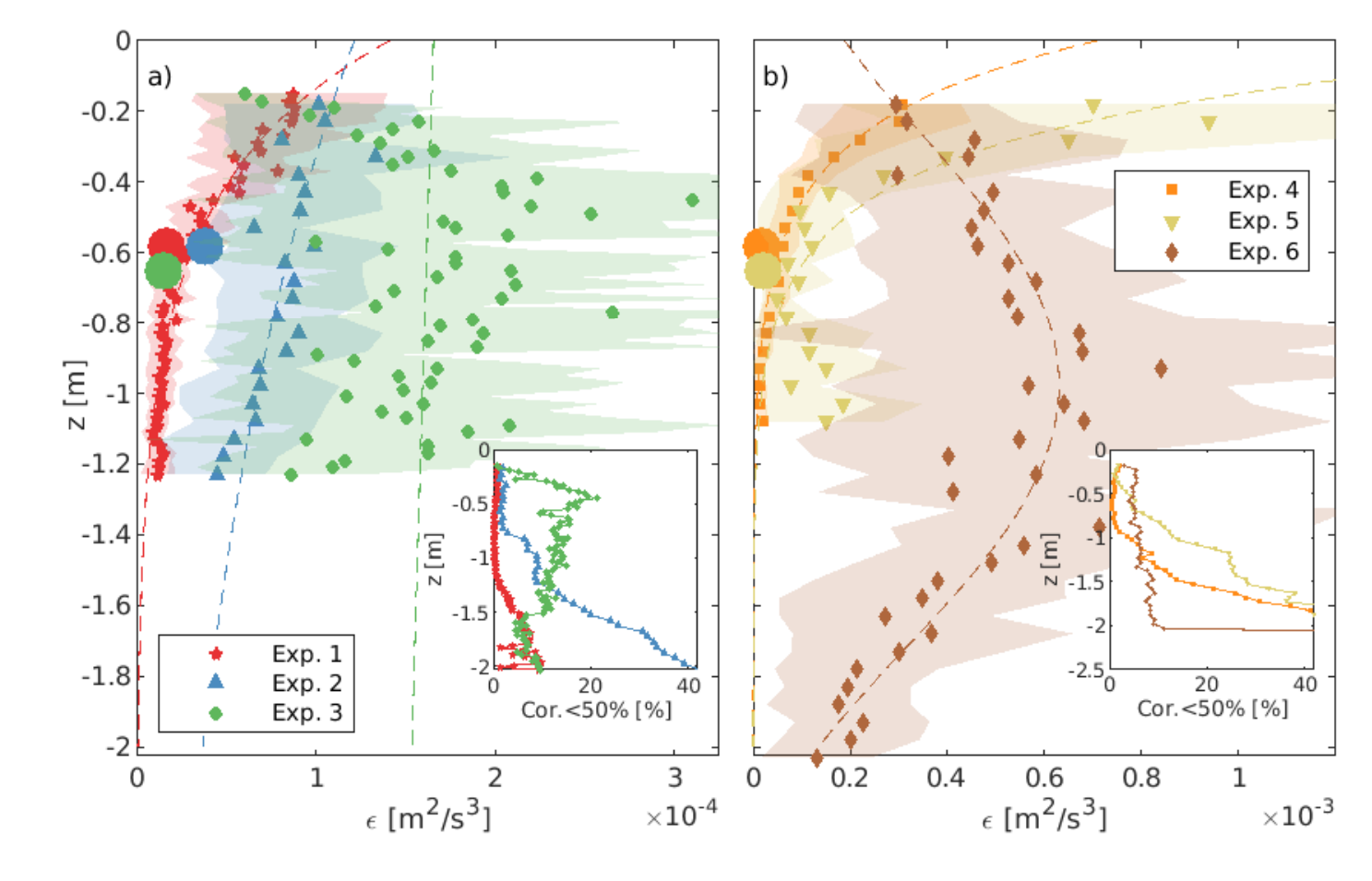}
    \caption{\label{fig:epsilon_profile} Estimated TKE dissipation rate profiles. a) ADCP placed 0.5~m from the pool edge in Exp.~1 (red), 2 (blue) and 3 (green). b) ADCP placed 0.25~m from the pool edge in Exp.~4 (orange) and 5 (yellow) and on the ice floe center in Exp.~6 (brown). In Exp.~1-5, $\epsilon$ was not estimated for $z<-1.2$~m due to the high instrument noise level. Dashed lines show curve fits to the ADCP data $\epsilon_{fit}$. Confidence intervals $\sigma_\epsilon$ are indicated with shaded regions. The inset plots show the vertical beam correlation data for the ADCP profiles (percentage of time series with correlation < 50\%). ADV data are presented as large dots.}
  \end{center}
\end{figure}

The spectra are proportional to $f^{-5/3}$ over a wide range of frequencies, meaning that both instruments were able to resolve the inertial subrange. Typically, ADCP data quality deteriorates as the distance from the instrument increases, either as decreasing beam correlation or increasing instrument noise. If the signal is obscured by Doppler noise, the spectra appear flat towards the higher frequencies. In figure~\ref{fig:spectra_profile}, the ADCP noise floor is in general $\sim10^{-5}~\mathrm{m}^{2}\mathrm{s}^{-1}$ close to the transducer and $\sim10^{-4}~\mathrm{m}^{2}\mathrm{s}^{-1}$ towards the end of the profile, with some exceptions, e.g. in Exp.~3 when the correlation was low (see figure~\ref{fig:epsilon_profile}~a). The ADV spectra exhibit a noise floor at $\sim10^{-6}~\mathrm{m}^{2}\mathrm{s}^{-1}$. In Exp.~1-5, the spectra from the bins below $z\approx-1.2$~m flatten out within the gray shaded region, which illustrates that the instrument noise level exceeded the TKE level for $f<f_2$. These data are not physical, hence not used to estimate $\epsilon$, which is only estimated for $z>-1.2$~m in Exp.~1-5 from Eq.~\ref{eq:epsilon}. However, all the spectra in Exp.~6 are approximately proportional to the $f^{-5/3}$ slope within the shaded region. Therefore, $\epsilon$ was estimated along the entire profile in Exp.~6.       

Figures~\ref{fig:epsilon_profile}~a-b present the estimated TKE dissipation rates $\epsilon$ from Exp.~1-3 and 4-6, respectively. The inset plots show the percentage of the ADCP time series where the vertical beam correlation was below the manufacturer recommendation (50\%). In Exp.~3, the beam correlation was below the recommended value more than 10\% of the time, see table~\ref{Table:stats}, which is an indication of poor data quality. This is probably why a large data scattering can be observed along the $\epsilon$ profile in Exp.~3. The profiles appeared to decay exponentially with depth when the ADCP was placed on the fast ice close to the pool wall, i.e. in Exp.~1-5, perhaps apart from Exp.~3 where the data quality was poor. Therefore, exponential functions on the shape $\epsilon_{fit}=ae^{-bz}$, where $a$ and $b$ are estimated parameters, were fitted to the data with nonlinear regression by means of iterative least squares and plotted as dashed lines in figure~\ref{fig:epsilon_profile}. A fourth order polynomial function was fitted to the estimated $\epsilon$ values in Exp.~6. The standard deviations of $\epsilon-\epsilon_{fit}$ along the profile are presented in table~\ref{Table:stats} as a measure on the accuracy of the curve fits. Especially the relative standard deviation, which is normalized over the mean $\epsilon$, shows that $\epsilon$ clearly decay exponentially with depth in Exp.~1, 2 and 4.    

\begin{table}[h]
\centering 
\begin{tabular}{c c c c c c c}  
\toprule
{Exp.} & {Corr. < 50\% [\%]} & {Mean $\epsilon$ [$\mathrm{m}^{2}\mathrm{s}^{-3}$]} & {Std. $\epsilon$ [$\mathrm{m}^{2}\mathrm{s}^{-3}$]} & {Rel. std. $\epsilon$ [\%]} & {$\pm \sigma_\epsilon$ [\%]} & {$d$ [$\mathrm{W}\mathrm{m}^{-2}$]} \\[0.5ex]
\midrule
1 & 0.5 & $3.4\times 10^{-5}$ & $6.1\times 10^{-6}$ & 17.8 & 49.6 & $5.7\times 10^{-2}$ \\[0.5ex]
2 & 4.3 & $8.3\times 10^{-5}$ & $1.2\times 10^{-5}$ & 15.0 & 49.4 & $1.5\times 10^{-1}$ \\[0.5ex]
3 & 12.7 & $1.6\times 10^{-4}$ & $4.7\times 10^{-5}$ & 28.9 & 63.5 & $3.3\times 10^{-1}$ \\[0.5ex]
4 & 2.3 & $8.4\times 10^{-5}$ & $1.5\times 10^{-5}$ & 18.4 & 45.6 & $1.7\times 10^{-1}$ \\[0.5ex]
5 & 8.1 & $2.4\times 10^{-4}$ & $1.3\times 10^{-4}$ & 53.1 & 71.1 & $4.4\times 10^{-1}$ \\[0.5ex]
6 & 5.3 & $5.2\times 10^{-4}$ & $8.6\times 10^{-5}$ & 16.6 & 66.4 & $9.2\times 10^{-1}$ \\[0.5ex]
\bottomrule
\
\end{tabular}
\caption{Statistics on ADCP beam correlation and estimated area density of TKE dissipation rate. Columns~2-6 apply for $z>-1.2$~m and column~7 applies for the entire profile, i.e. $z>-2.0$~m.  Column~2 is the percentage of the time series where the beam correlation was less than 50\%, averaged over all bins. Column~3 is the mean $\epsilon$ averaged over all bins. Column~4 is the standard deviation of $\epsilon-\epsilon_{fit}$. Column~5 is the relative standard deviation, i.e. column~4/column~3. Column~6 is the average uncertainty in the estimated $\epsilon$, given in Eq.~\ref{eq:epsilon_sigma} and the shaded area in figure~\ref{fig:epsilon_profile}. Column~7 is the area density of TKE dissipation rate from the $\epsilon_{fit}$ profiles.} 
\label{Table:stats} 
\end{table} 

It is desirable to quantify the total energy dissipated in turbulence in the water affected by the ice floe motion. The $\epsilon_{fit}$ values were therefore numerically integrated with the trapezoidal method over the entire profile to find the area density of TKE dissipation rate $d=\rho_w \int_{-2}^{0}\epsilon_{fit} \,dz~[\mathrm{W}\mathrm{m}^{-2}]$. Confidence intervals for the estimated $\epsilon$, i.e. $\sigma_\epsilon$ estimated from Eq.~\ref{eq:epsilon_sigma}, are illustrated as shaded regions in figure~\ref{fig:epsilon_profile}. Estimated $d$ and the average percentage of the uncertainties with respect to the fitted curves are listed in table~\ref{Table:stats}. Estimated $\epsilon$ from the ADV spectra are presented as large dots in figure~\ref{fig:epsilon_profile}. Some of the dots are displaced a bit in the vertical direction to increase the readability, even though the measurement volume was located at $z=-0.58$~m in all the experiments. The estimated values from the ADV were in general smaller than the values from the ADCP. The reason for this is unknown in Exp.~1-3 but is probably that the ADCP was placed closer to the pool, where the TKE level is expected to be higher, in Exp.~4-5. Ideally, the ADV should have been mounted at the same $x$-position as the ADCP, but this was not possible with the ADV tripod. 

\begin{figure}
  \begin{center}
    \includegraphics[width=.75\textwidth]{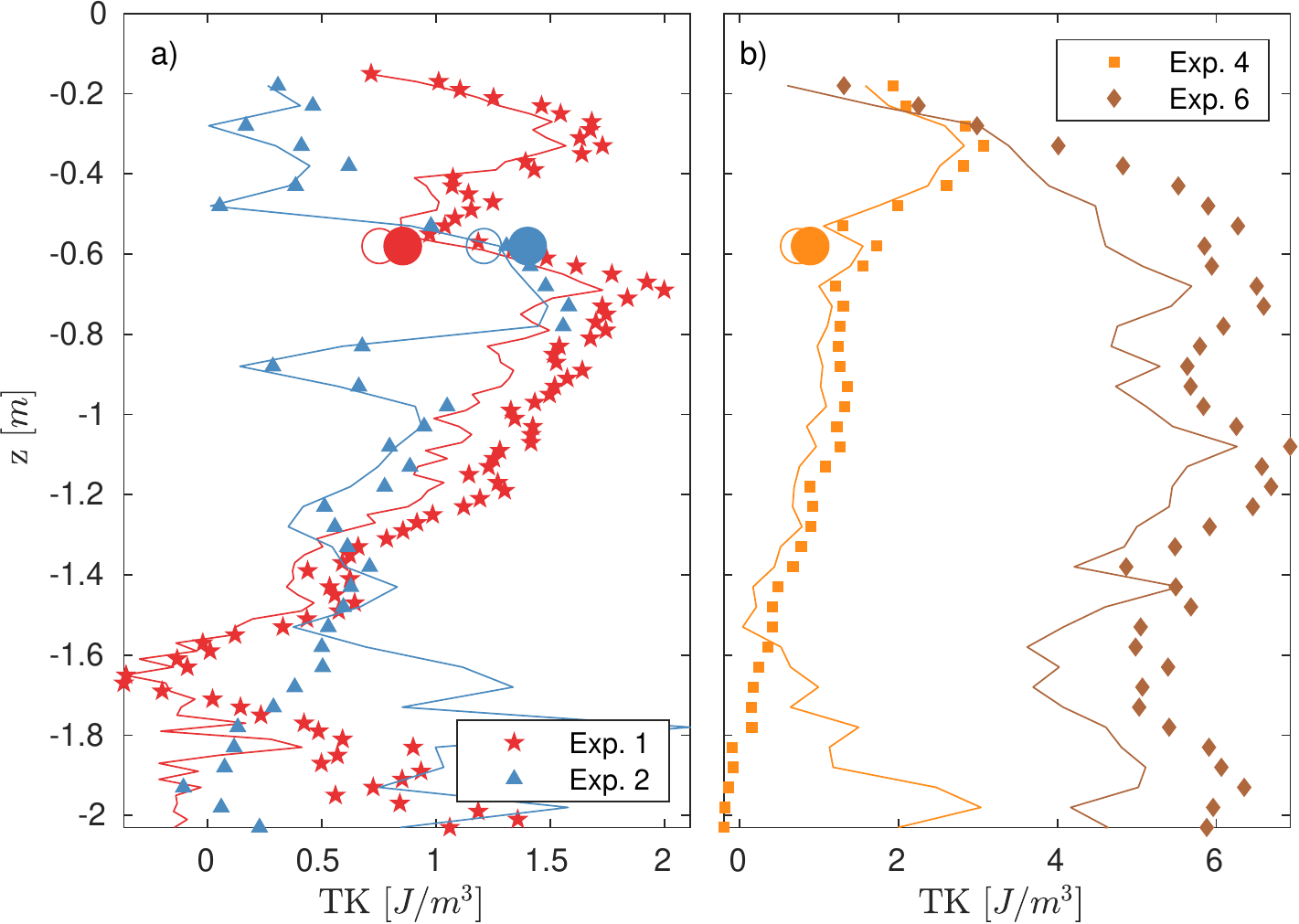}
    \caption{\label{fig:TKE_profile} Estimated density of TKE profiles. a) ADCP placed 0.5~m from the pool edge in Exp.~1 (red) and 2 (blue). b) ADCP placed 0.25~m from the pool edge in Exp.~4 (orange) and on the ice floe center in Exp.~6 (brown). Experiments~3 and 5 showed the same behavior as Exp.~1-2 and 4, respectively, but these are not included to increase readability. ADV data are presented as large dots. Solid markers show measured values and lines show values corrected for instrument noise.}
  \end{center}
\end{figure}

The density of TKE profiles $TK$ from the ADCP and single values from the ADV obtained from Eqs.~\ref{eq:TKE_ADCP}-\ref{eq:TKE_ADV} are presented in figure~\ref{fig:TKE_profile}, where solid markers indicate measured data and solid lines indicate data corrected for instrument noise. There is good agreement between the ADCP and the ADV, especially in Exp.~1-3 where the instruments were placed at the same $x$-location. As previously discussed, it is expected that the TKE level was higher closer to the pool edge, which probably explains the lower values obtained from the ADV in Exp.~4-5. The density of TKE profiles approach zero with increasing depth and are therefore numerically integrated over the profile to find the area density of TKE $TK_z= \int_{z_2}^{z_1} TK \,dz~[\mathrm{J}\mathrm{m}^{-2}]$, where $z_{1}$ is the first ADCP bin and $z_{2}$ is the last considered bin. The profiles show some negative values and other unphysical behavior in depths below $z_{2}$, which is set to -1.2 and -2~m in Exp.~1-5 and 6, respectively, in consistency with Figs.~\ref{fig:spectra_profile}-\ref{fig:epsilon_profile}. The $TK_z$ values are listed in table~\ref{Table:current}.

Turbulence properties from the tidal current were investigated to find the ambient turbulence level in the boundary layer below the ice. Reference runs of ADCP time series before each experiment, i.e. when the ice floe was not moving, were considered. The mean horizontal current speed $U_{mean}$ and direction $U_{dir}$, averaged over bins above $z=-1.2$~m, as well as the duration of the reference runs, are summarized in table~\ref{Table:current}. The direction is defined as clockwise rotation about the $x$-axis and the mean current direction was approximately in the $y$ and $-y$-direction in Exp.~1-5 and 6, respectively. The ADCP was usually started right before the experiments, hence the short reference run time series. 

Only in Exp.~1, the reference run was long enough to estimate the ambient area density of TKE dissipation rate $d_{amb}$ and the ambient area density of TKE $TK_{z,amb}$. However, $U_{mean}$ was quite consistent in the order of $10^{-2} \mathrm{m}\mathrm{s}^{-1}$ over Exp.~1-5, so it is reasonable to assume that $d_{amb}$ and $TK_{z,amb}$ were similar in these experiments. In Exp.~6 on the other hand, $U_{mean}$ was smaller, in the order of $10^{-3} \mathrm{m}\mathrm{s}^{-1}$. Nevertheless, the values of $d_{amb}$ and $TK_{z,amb}$ obtained in Exp.~1 are used as conservative estimates for all the experiments. Both parameters are resulting from the tidal current and are listed in table~\ref{Table:current}. Depending on the experiment and the location of the ADCP, $d_{amb}$ was 1.4-22.8\% of $d$, and $TK_{z,amb}$ was only 0.8-8.1\% of $TK_z$. In the following, $d_{amb}$ and $TK_{z,amb}$ due to the mean tidal current are subtracted from $d$ and $TK_z$, respectively, which contain TKE from the moving floe and the tidal current, so that $d=d-d_{amb}$ and $TK_{z}=TK_{z}-TK_{z,amb}$. Henceforth, focus is directed towards the TKE dissipation rate due to the moving floe.     

\begin{table}[h]
\centering 
\begin{tabular}{c c c c c c c}  
\toprule
{Exp.} & {$U_{mean}$ [mm/s]} & {$U_{dir}$ [$^{\circ}$]} & {Time [s]} & {$d_{amb}$ [$\mathrm{W}\mathrm{m}^{-2}$]} & {$TK_{z}$ [$\mathrm{J}\mathrm{m}^{-2}$]} & {$TK_{z,amb}$ [$\mathrm{J}\mathrm{m}^{-2}$]} \\[0.5ex]
\midrule
1 & 7.9 & 253 & 485 & $1.3\times 10^{-2}$ & 1.3 & $6.5\times 10^{-2}$ \\[0.5ex]
2 & 4.5 & 253 & 75 & - & 0.8 & - \\[0.5ex]
3 & 8.0 & 234 & 175 & - & 1.2 & - \\[0.5ex]
4 & 5.4 & 267 & 170 & - & 1.5 & - \\[0.5ex]
5 & 5.4 & 267 & 170 & - & 0.9 & - \\[0.5ex]
6 & 1.4 & 104 & 190 & - & 8.4 & - \\[0.5ex]
\bottomrule
\
\end{tabular}
\caption{Ambient flow with mean horizontal current and direction and length of the reference run time series (columns~2-4). Experiment~4 and 5 were conducted within one hour and it is assumed that the tidal conditions were similar. Column~5 is the ambient area density of TKE dissipation rate. Area density of TKE during towing and due to the tidal current are listed in Column~6 and 7, respectively. $d_{amb}$ and $TK_{z,amb}$ were only estimated in Exp.~1 due to sufficient duration of the reference run time series. } 
\label{Table:current} 
\end{table}            

The total TKE dissipation rate $D=d S_{b0}$, where $S_{b0}$ is the horizontal area of the pool, describes the rate of TKE dissipation due to the ice floe motion in the water volume below $S_{b0}$, and is analogous to the first integral in Eq.~\ref{eq:D_water}. The total TKE advection rate $TK_{adv} = TK U_{mean} S_{l0}$ describes the rate of TKE due to the floe motion that is transported away from the water volume below $S_{b0}$ due to the mean current speed $U_{mean}$ and dissipated elsewhere, and is analogous to the second integral in Eq.~\ref{eq:D_water}. $S_{l0}$ is the area of the vertical, cylindrical surface separating the pool from the fast ice, projected on a plane with normal vector parallel to $U_{dir}$. As the mean horizontal current direction was roughly parallel to the $y$-axis, $S_{l0}$ is approximately parallel to the $xz$-plane. Since $TK$ is already integrated over $z$ to obtain $TK_{z}$, $TK_{adv} = TK_z U_{mean} L_{l0}$, where $L_{l0}$ is the length of $S_{l0}$ in the $x$-direction, i.e. $L_{l0} \approx 6~\mathrm{m}$.  

In order to accurately quantify the total TKE dissipation due to the moving ice floe, the ADCP should have been deployed at many locations around the pool and on the ice floe, so that $D$ and $TK_{adv}$ could have been estimated with a high spatial resolution in the horizontal plane. An attempt is still made to estimate the total TKE dissipation rate $D$ and the total TKE advection rate $TK_{adv}$. From figure~\ref{fig:epsilon_profile}, it is clear that the profiles of TKE dissipation rate are very different in the gap between the floe and the fast ice, where $\epsilon$ decay exponentially with depth, and below the floe itself, where $\epsilon$ first increase and then decay with depth after a maximum is reached at $z\approx-1$~m. This difference is also apparent for the profiles of the density of TKE in figure~\ref{fig:TKE_profile}. Therefore, the representative area and length are separated so that $D = d_fS_f+d_{gap}S_{gap}$ and $TK_{adv} =  TK_{z,f} U_{mean} L_{f} + TK_{z,gap} U_{mean} L_{gap}$, where the notation \textit{f} indicates the horizontal area and length of the ice floe, and \textit{gap} indicates the horizontal gap area and length in the $x$-direction. 

It is assumed that the $\epsilon$ and $TK$ profiles obtained in Exp.~6 are representative for the TKE below the entire ice floe, hence $S_{f}=12~\mathrm{m}^{2}$, $L_{f}=4~\mathrm{m}$, $d_{f}=d_6$ and $TK_{adv,f} = TK_{adv,6}$. From Figs.~\ref{fig:plume_evolution}-\ref{fig:plume_PIV}, it can be observed that the jet diameter (and the resulting turbulent cloud) is $\sim 1~\mathrm{m}$, and it is assumed that the jet extension in the $y$-direction is equal to the width of the floe $W_f$ and that a similar jet is produced in the gap on the north end of the floe, hence $S_{gap}=6~\mathrm{m}^{2}$ and $L_{gap}=2~\mathrm{m}$, which is in agreement with the total gap area and length in both short ends of the pool. As discussed in Section~\ref{subsec:ROV_jet}, the ADCP data obtained at the shortest distance from the pool, i.e. in Exp.~4-5, are probably a better realization of the flow happening in the gap than further away from the pool. Out of these two, the most correlated beam measurements, the least uncertainties in the estimated $\epsilon$ and the most cycles were obtained in Exp.~4. Therefore, it is assumed that the $\epsilon$ and $TK$ profiles obtained in Exp.~4 are representative for the TKE in the entire gap area, hence $d_{gap}=d_4$ and $TK_{adv,gap} = TK_{adv,4}$. With these assumptions, $D=11.8~\mathrm{W}$, where the weighted uncertainty from $\sigma_\epsilon$ is $\pm64.6$\%. Similarly, $TK_{adv}=0.06~\mathrm{W}$. The total TKE rate due to the floe motion $D+TK_{adv}=11.9~\mathrm{W}$, which corresponds to Eq.~\ref{eq:D_water}, is estimated to be 36.9\% of the input power $P_{winch}$.

\section{Discussion} \label{sec:Discussion}

As the ice floe was towed back and forth in the pool, an oscillating flow was generated in the surrounding water due to the shear at the water-ice interface. This large-scale water flow is considered turbulent in itself because it was fluctuating about a mean value (close to zero), as seen in figure~\ref{fig:timeseries_AD}. In the TKE spectra shown in figure~\ref{fig:spectra_profile}, these large-scale fluctuations appear around the peak frequency of 0.04~Hz, corresponding to periods around 26~s, i.e. the mean duration of a cycle. Two different mechanisms generated turbulence in the pool, the drag associated with the relative water-ice velocity, and the downward jet injection and upward suction of fluid in the gap. The former creates a turbulent boundary layer below the oscillating floe, which can be observed in the TKE dissipation rate $\epsilon$ profile in Exp.~6 presented in figure~\ref{fig:epsilon_profile}. This is likely to occur around natural ice floes due to wave induced motion of water particles relative to the ice, and comprises turbulent friction on the underside of the ice and the wake behind the sharp edges of the floe, i.e. skin friction and form drag, respectively \citep{kohout2011wave}. The latter induces turbulence that rapidly decays with depth, as seen in the $\epsilon$ profiles in Exp.~1-5, which also may occur in a dense floe field exposed to waves \citep{rabault2019experiments}. Note that the epsilon profiles in Exp.~1-5 and 6 comprise turbulence from both the jet and suction motion and towing back and forth, respectively, as the entire time series including all cycles were used. 

Under the assumption that the entire energy transfer from the ice floe to the water was dissipated in turbulence, either directly below the system or advected away from the system with the mean horizontal current, the estimated total TKE rate due to the floe motion $D+TK_{adv}=11.9~\mathrm{W}$ presented in Section~\ref{subsec:TKE_energy} corresponds well to the calculated power of the ice floe work to move the surrounding water $P_{drag}=11.6~\mathrm{W}$ presented in Section~\ref{sec:theoretical_background}. Note that these estimates contain large uncertainties, such as the friction coefficients used in the calculations and other aspects discussed in the next paragraph in the case of the measurements.         


It was found that 36.9\% of the mean measured power input to the system $P_{winch}$ was dissipated in turbulence arising from water-ice friction, although a large uncertainty is associated with this estimate, and it should be used with caution. More than 80\% of the total TKE rate due to the floe motion occurred under the ice floe, based on the information acquired in Exp.~6, which is associated with the relative water-ice velocity and floe drag. The data quality was good in this case, but the statistical confidence is reduced due to the fact that this experimental setup was only repeated once. In addition, the average uncertainty in the estimated TKE dissipation rate $\sigma_{\epsilon,6}$ was 66.4\%. Due to the lack of further measurements, it was assumed that the area density of TKE dissipation rate $d$ was uniform over the area of the floe, which is probably a large simplification of reality. The $\epsilon$ profiles in Exp.~1-5, shown in figure~\ref{fig:epsilon_profile}, are qualitatively consistent in the sense that they all, perhaps apart from Exp.~3, decay exponentially with depth. This behavior is attributed to the downward water jets and upward suction motions in the short ends of the pool. However, there are quantitatively large discrepancies in these estimated $\epsilon$ profiles, expressed through $d$. For example, $d_5$ was 2.6 times greater than $d_4$, even though the experimental setups were identical. Experiment~4 contained more cycles, higher data quality and lower $\sigma_{\epsilon}$ (45.6\%), hence $d_4$ was used to estimate the total TKE dissipation rate $D_{gap}$. Note that the turbulence induced by the shear flow in the gap between the fast ice and the lateral sides of the floe was not measured and has not been accounted for. 

\begin{figure}
  \begin{center}
    \includegraphics[width=.6\textwidth]{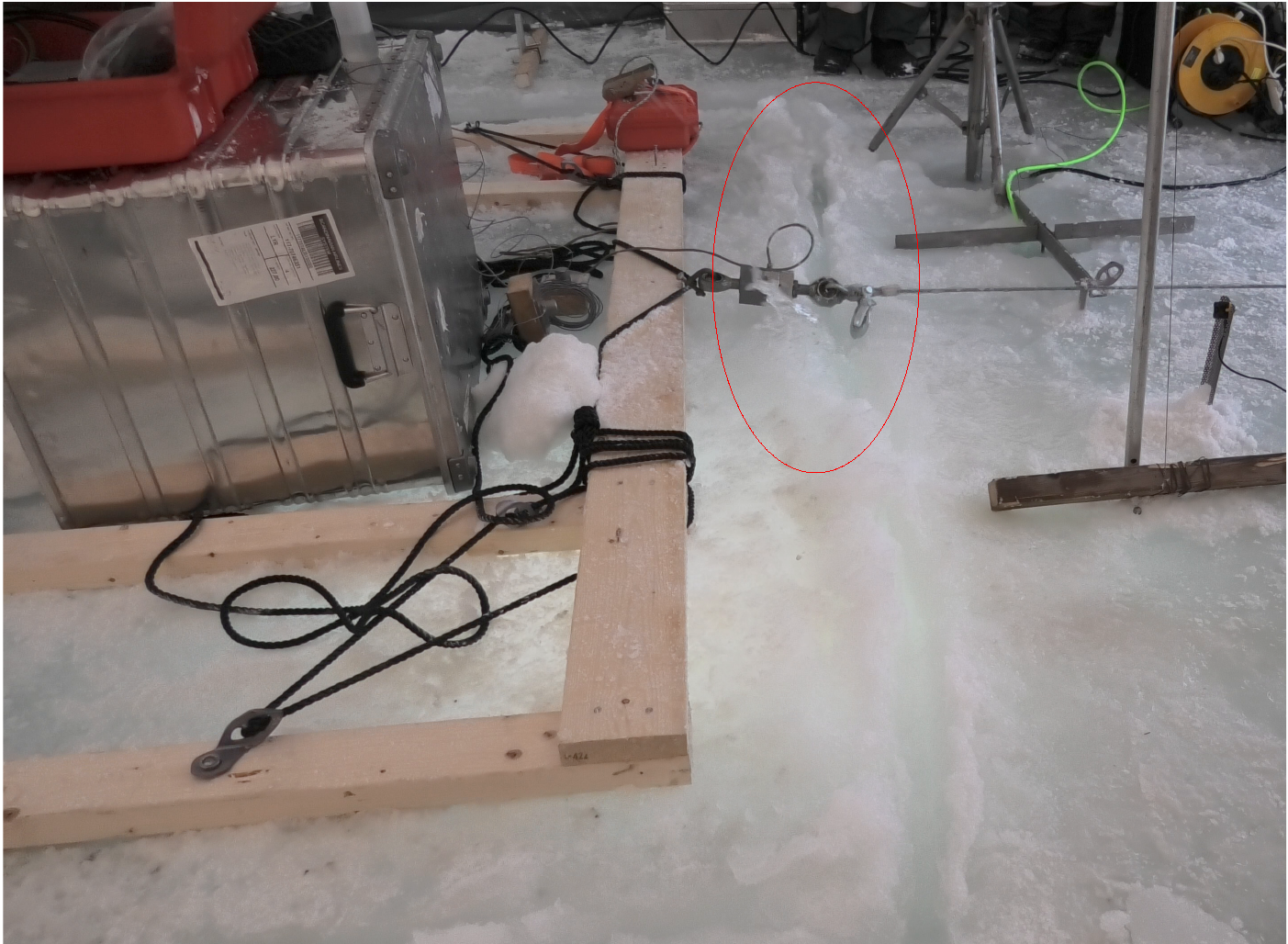}
    \caption{\label{fig:splash} Owerwash immediately after a collision event. The red circle marks the erupting water jet. }
  \end{center}
\end{figure}

Approximately 7.5\% of the of the mean power input $P_{winch}$ was on average absorbed in the collisions between the ice floe and the pool walls $P_{coll}$, probably in mechanisms such as inelastic deformation of the ice and erosion/slush production \citep{herman2018wave}. Figure~\ref{fig:splash} shows an image of a collision event where the walls of impact have been deformed and slush is building up on the topside of the ice. A central question is whether collisions of similar characteristics take place between natural floes in a wave field. Several studies agree with the present findings. \cite{martin1987high,martin1988ice} investigated large ice floes in the order of $10^{2}$~m heavily concentrated in the Greenland and Bering Sea MIZ and found that collision events in general recurred with the period of the ocean swell, which was 10-18~s, i.e. a bit shorter than the 26~s period used in the present experiments. They also found that the collision energy was correlated with, and about 10\% of the total wave energy. \cite{shen1998wave} found from modeling that energy absorption for typical ocean swell periods arising from collisions between adjacent ice floes in the order of 1~m in a dense pancake ice field is the second most dominating mechanism in terms of energy dissipation, after TKE in the water column. \cite{li2018laboratory} presented wave tank experiments with ice floes in the order of 1~m, which suggested that approximately 10\% of the wave energy was dissipated in inelastic collisions between adjacent floes.  

Induced TKE and inelastic collisions were estimated to dissipate approximately 45\% $\pm$23.7\% of the total input energy rate, hence these mechanisms do not account for all dissipation processes. Another possible loss is the generation of outgoing surface waves in the pool due to the floe motion, which is directly associated with the damping force of the body \citep{squire1995ocean}. Surface waves were observed visually in the pool and may be associated with the 0.5~Hz oscillations in the uniaxial accelerometer time series presented in the lower panel of figure~\ref{fig:timeseries_evo}. The frequency of the piston mode wave oscillations $f_p$ in an oscillating water column can be described by $2\pi f_p = \sqrt{g/h}$, where $h$ is the height of the water column \citep{baudry2013overview}. When $H_f$ is substituted with $h$, $f_p = 0.5$~Hz is obtained, which agrees with the frequency of the measured floe oscillations. Floe generated surface waves have also been observed in real ice-wave fields. According to \cite{wadhams1988attenuation}, the nonlinear energy transfer from the long swells via ice floes to shorter waves due to floe motion and interactions could be a possible explanation for the rollover effect, i.e. a generation of short period waves deep into the MIZ, where these waves are expected to be damped by the presence of the ice. Overwash or water jets were also observed as a consequence of the collisions, which is another damping mechanism that may influence the attenuation of surface waves in a wave-ice field \citep{herman2018wave,marchenko2019influence}. An example of a splashing event is shown in figure~\ref{fig:splash}. Some energy may also have dissipated in the towline and ice screws. None of the above-mentioned mechanisms were measured, only observed, and are therefore not quantified in the present study. 

It is assumed that the loss/absorption due to hysteresis caused by flexural floe deformation is low due to the high aspect ratio of the floe, hence negligible bending motion \citep{meylan1994response}. This assumption agrees with the findings of \cite{shen1998wave}. Losses caused by floe bending motion would be more relevant for floes with smaller aspect ratio exposed to waves with wavelength comparable to the diameter \citep{meylan1994response}.

Processes of energy dissipation around moving ice floes are of importance for numerical forecasting on wave attenuation in the Arctic and climate models including momentum flux in the atmosphere-ice-ocean system and upper ocean mixing. Direct observations, such as the present study, need to be generic and transferable to a distribution of floe sizes to be applicable in a modeling perspective. The ice floe dimensions used here may be realistic for floes in the outer part of the MIZ where the dominating short waves break up the ice layer into small floes. Further into the MIZ, the floe thickness may still be in the order of 1~m, but a typical floe length $L_f$ is one order of magnitude larger than the present floe. The relative water-ice velocity could be similar to what was obtained in the pool. With these assumptions, a discussion follows on the scaling of different contributions to power transfer to the water with respect to the floe length, based on the mathematical formulations in Section~\ref{sec:theoretical_background}. Form drag power and jet power increase linearly to the floe length $P_{fd},P_{jet} \propto L_f$. This applies also for skin friction power on the lateral floe sides, while skin friction power on the floe bottom increases quadratically to the floe length $P_{sd} \propto L_f^2$. The floe mass and thus kinetic energy will also increase quadratically to the floe length $K_{f} \propto L_f^2$. The power transfer due to floe-floe collisions is related to the kinetic energy of the floe, but may also be related to the contact area, which scales proportionally to the floe length. Note that in the current experimental setup, collisions were present in each towing cycle, which may occur under some wave and wind conditions, but ice floe fields can also disperse under other conditions.            


\section{Conclusions} \label{sec:Conclusions}

Various mechanisms of energy dissipation and floe dynamics around a colliding full-scale ice floe have been investigated experimentally in an Arctic environment, and the paper presents much needed direct turbulence measurements. Wave induced motion was simulated by towing the floe in an artificially made pool in the fast ice, back and forth in an oscillatory manner so that collisions with the fast ice occurred. The constructed setup corresponds to a dense field of small sized ice floes responding in surge when acted upon by long period ocean swell typically found in the MIZ. Extensive instrumentation, i.e. a load cell, a range meter, accelerometers, an IMU, an ADV, a high-resolution ADCP and an ROV, allowed for detailed surveillance of the towing load, floe motion and kinematics of the surrounding water. The average rate of input energy to the system, found from the towing load and the floe translational velocity in the axial direction, was 32.2~W. 

Turbulence was generated from the relative water-ice velocity, comprising turbulent friction on the underside of the ice and the wake behind the floe, and from the downward water jets and upward suction motion associated with the collision events. The latter phenomenon was visualized with a new technique with rising bubbles and an ROV as tracers and camera, respectively. TKE frequency spectra were found to contain an inertial subrange where energy was cascading at a rate proportional to $f^{-5/3}$, according to Kolmogorov's theory. From spectral analysis, the total TKE rate due to the floe motion was estimated to be 11.9~W $\pm$64.6\%, which corresponds to 36.9\% $\pm$23.7\% of the input energy rate. The dominating mechanism for wave energy dissipation in ice floe fields is still debated. Despite relatively high uncertainties, these results indicate that a substantial portion of the wave energy is dissipated in turbulence. From the accelerometer data, energy absorption due to collisions was calculated as the change in the kinetic energy of the floe immediately before and after the collision events. The estimated rate of energy loss in this process was 2.4~W, i.e. 7.5\% of the input energy rate, which was attributed to inelastic ice deformation and slush production.

\section*{Acknowledgments}

The authors are grateful to Andrej Sliusarenko, Vladimir Markov and Sergej Podleshych for their assistance. We also thank Olav Gundersen for the winch setup and Store Norske for accommodating the researcher in Svea while performing the experiments. Funding for the experiment was provided by the Research Council of Norway under the PETROMAKS2 scheme (DOFI, Grant number 28062) and the IntPart project Arctic Offshore and Coastal Engineering in Changing Climate (Project number 274951). The data are available from the corresponding author upon request.

\section*{Appendix A - Instrument synchronization}

It was necessary to synchronize the range meter and load cell time series in the post processing in order to calculate the winch power applied on the ice floe. The synchronization scheme described herein applies for the instruments marked with diamonds in table~\ref{Table:instruments}. As mentioned in Section~\ref{subsec:Instrumentation}, the range meter was sampled in correct UTC time in the first place. The load cell and the accelerometers were connected to the same data acquisition unit and therefore synchronized, but the computer clock was off. The data were first re-sampled to a common sampling rate of 1~kHz. From the uniaxial accelerometer time series, distinct peaks were recognized at the instance of impact, as elaborated in Section~\ref{subsec:Collision_energy}. The IMU, which sampled in correct GPS time, produced the same peaks in the time series. Hence, the correct UTC time of the first impact in Exp.~3 was found from the IMU time series. Finally, the acceleration and load cell time series were shifted to coincide with this instance.

\bibliographystyle{agsm}
\bibliography{template_Svea.bib}

\end{document}